\RequirePackage{fix-cm}
\documentclass[smallextended]{svjour3}       % onecolumn (second format)
\smartqed  % flush right qed marks, e.g. at end of proof
\usepackage[]{url}
\usepackage{graphicx}
\usepackage{comment}
\usepackage{multirow}
\usepackage{color}
\usepackage{lscape}
\usepackage{longtable}
\usepackage{supertabular}
\usepackage{subfig}
\usepackage{graphicx}
\usepackage{natbib}
\usepackage[table]{xcolor}% http://ctan.org/pkg/xcolor
\usepackage[nomessages]{fp}% http://ctan.org/pkg/fp
\newlength{\maxlen}

\newcommand{\rqtwo}{\textit{What SE topics relate to SE human capital?}\newline}
\newcommand{\rqthree}{\textit{What theories have been analyzed for SE human capital studies?}\newline}
\newcommand{\rqfour}{\textit{Where does the data originate from for studies related to SE human capital?}\newline}
\newcommand{\sayaOSS}{SE-HCI}
\newcommand\myworries[1]{\textcolor{red}{#1}}

\begin{document}

\title{Human Capital in Software Engineering: 
A Systematic Mapping of Reconceptualized \\ Human Aspect Studies}
%\title{How Do Humans Capital in Software Development? A Systematic Mapping}
%\subtitle{Do you have a subtitle?\\ If so, write it here}

\titlerunning{Human Capital in Software Engineering}        % if too long for running head

\author{Saya Onoue \and
        Hideaki Hata \and
        Raula Gaikovina Kula \and
        Kenichi Matsumoto
}

%\authorrunning{Short form of author list} % if too long for running head

\institute{Saya Onoue, Hideaki Hata, Raula Gaikovina Kula and Kenichi Matsumoto \at
              Nara Institute of Science and Technology, Japan \\
              \email{\{onoue.saya.og0, hata, raula-k, matumoto\}@is.naist.jp}           %  \\
%             \emph{Present address:} of F. Author  %  if needed
}

\date{Received: date / Accepted: date}
% The correct dates will be entered by the editor

\maketitle

\begin{abstract}
The human capital invested into software development plays a vital role in the success of any software project.
By \textit{human capital}, we do not mean the individuals themselves, but involves the range of knowledge and skills (i.e., human aspects) invested to create value during development.
However, there is still no consensus on how these broad terms of human aspects relate to the health of a project. 
In this study, we reconceptualize human aspects of software engineering (SE) into a framework (i.e., SE human capital).
The study presents a systematic mapping to survey and classify existing human aspect studies into four dimensions of the framework: capacity, deployment, development, and know-how (based on the Global Human Capital Index).
%proposed and reported by the World Economic Forum).
From premium SE publishing venues (five journal articles and four conferences), we extract 2,698 hits of papers published between 2013 to 2017.
Using a search criteria, we then narrow our results to 340 papers.
Finally, we use inclusion and exclusion criteria to manually select 78 papers (49 quantitative and 29 qualitative studies). 
Using research questions, we uncover related topics, theories and data origins.
The key outcome of this paper is %extraction of existing metrics that are 
a set of \textit{indicators} for SE human capital. 
This work is towards the creation of a SE Human Capital Index (\sayaOSS) to capture and rank human aspects, with the potential to assess progress within projects, and point to opportunities for cross-project learning and exchange across software projects.

% \PACS{PACS code1 \and PACS code2 \and more}
% \subclass{MSC code1 \and MSC code2 \and more}
\end{abstract}

\section{Introduction}
\label{intro}

Software development involves a variety of human-intensive activities.
Researchers refer to it as knowledge-intensive \citep{WOHLIN-2015}, with developers (i.e., humans) playing an important role the success of a software project \citep{DeMarc-2013}.
As an intangible asset, software development produces artifacts that lack physical substance (unlike physical assets such as machinery and buildings).
%Rather than 
Therefore, simply evaluating any software system in terms of the physical humans is not practical, as it involves a much more deeper investment of human activities related to the skills and knowledge that is created, retained and lost during development.
The numerous metrics and frameworks defined to measure health is evidence of the difficultly when assessing the health of an OSS project \citep{Crowston-ieee2006}. 
To date, there is no single consensus on the broad terms related to human aspects in software engineering.
%In recent times, research such as Wohlin et. al., provided an adoption of economic perspectives such as intellectual capital (a superset of human capital) into software engineering.

In economics, human investments are seen as capital -- specifically intellectual capital, which is \textit{``the sum of all knowledge firms utilize for competitive advantage''} \citep{Nahapiet1998,EconBiz-2004,WOHLIN-2015} and can be characterized into different forms (e.g., human, social, relational, structural, etc)\footnote{Capital covers people (i.e., human capital), the value inherent in its relationships (i.e., relational capital), and everything that is left when the employees go home (i.e., structural capital).}.
Oxford handbook on Human Capital highlights the importance of human capital, that is, \textit{``all forms of intellectual capital including social and structural capital are arguably reducible to the human knower, thus human capital becoming the linchpin''}~\citep{RePEc:oxp:obooks:9780199655892}.
In software engineering, Wohlin et. al. argues that intellectual capital is the bigger umbrella, hence defined human capital as a combined form of  social and organization capital.
%Similarly, prior studies focused on human activity metrics, including productivity, communication structure, or knowledge loss.
Similarly, we do not distinguish other forms of intellectual capital, but instead use human-aspects to map into human capital.
%We revisit the original concepts, making a more in-depth analysis of \textit{human capital} in software development. %in terms of its four dimensions.

%The authors used a bigger umbrella, by which they propose that human capital is paired with social and organization capital \cite{WOHLIN-2015}.

%By leveraging concepts of the economic viewpoint of `Human Capital', we can closely analyze and empirically measure various dimensions of the Human Capital into software development.
%In detail, the ``skill, knowledge and similar attributes that affect particular human capabilities to do productive work" which can be improved through health facilities, on-the-job training, formal education and study programmed \cite{Schultz-1961}. 
%This capital resides with, and is utilized by individuals.

% \myworries{HC, GHCI}
A practical implementation of human capital is the Global Human Capital Index (GHCI)\footnote{http://reports.weforum.org/global-human-capital-report-2017/}, as proposed and reported by the World Economic Forum\footnote{https://www.weforum.org}.
In 2017, the GHCI 2017 ranked 130 countries on how well they are developing their human capital on a scale from 0 (worst) to 100 (best) across four thematic dimensions and five distinct age groups to capture the full human capital potential profile of a country.
Our intention is to reconceptualize human aspects in software engineering (SE), hence, building a framework of software engineering human capital (SE human capital) which is similar to GHCI.

% \myworries{SMS}
% A systematic mapping study is a type of systematic literature review. 
% A systematic literature review is a repeatable method for identifying relevant studies to answer specific research questions \cite{Mendes-tse2010}. 
% In particular, a systematic mapping study is designed to give an overview of a research area through classification and counting contributions in relation to the categories of that classification \cite{Kitchenham2007, Petersen2008,Petersen-ist2015}.
% Kitchenham et al. \cite{Kitchenham-ease2010} contrasted the different characteristics of the process of systematic literature reviews and mapping studies. 
% A systematic mapping study follows the same principled process as systematic literature reviews, though there are different criteria for inclusions/exclusions and quality \cite{Wohlin2000}. 
% Systematic mapping studies do not have strict roles compared to systematic literature reviews, however, various types of papers should be reviewed for understanding target
% topic area widely.

This paper is an investigation into how human aspects can be reconceptualized into SE human capital.
%The study involves a systematic mapping to survey and classify existing human-aspect papers. 
%Our main goal is to identify what existing metrics describe human capital in SE.
We carried out a systematic mapping from five top journal articles and four top international conferences published between 2013 to 2017. 
Based on the GHCI framework, we then classified the 78 studies into four dimensions: \textit{capacity} for skill attainment, \textit{deployment} for a workforce, \textit{development} for upskilling and reskilling, and \textit{know-how} for specialized skills.
The mapping study addresses three areas of SE human capital, which is (a) topics, (b) theories and (c) data origins.
%Results indicate that topics ranged (i.e., Capacity: (1a) \textit{personality}, (1b) \textit{success}, and (1c) \textit{performance}, 2. Deployment: (2a) \textit{OSS} (2b) \textit{community}, (2c) \textit{communication} and (2d) \textit{productivity}, 3. Development: (3a) \textit{user} and (3b) \textit{OSS} and 4. Know-how: (4a) \textit{practices}).
Results show that there are specific topics in each dimension. 
%(i.e., \textit{personality}, \textit{success}, and \textit{performance} for capacity, \textit{OSS}, \textit{community}, \textit{communication}, and \textit{productivity} for deployment, \textit{user} and \textit{OSS} for development, and \textit{practices} for know-how).
%Interestingly,
Regarding theories, we find that %77\% of papers do not report any theory and the experiment research method is a most common type of research.
18 papers described theories in their studies: 10 theories are identified including game theory, organization theory, signaling theory, and so on.
In terms of the data, half of the papers tend to use multiple sources of data, often combining code and other assets. 
We also found that %58\% of papers analyzed less than 10 projects with them data ranged more than five years.
most papers studied OSS projects (83\%), although some analyzed company data.

Our main contribution is a listing of \textit{indicators} derived from summarizing the mapped studies.
This listing is useful for constructing a SE Human Capital Index (\sayaOSS) in the future.
Much like the GHCI, the \sayaOSS~has potential to reveal insights into how projects develop their human capital and will be used as an important determinant of their long-term success.
%than virtually any other factor.
%We envision that HCI as a ranking of software projects to capture the full human capital potential profile.
%It aims to be used as a tool to assess progress within projects and point to opportunities for cross-country learning and exchange across projects.

The rest of the paper is outlined as follows. %Section \ref{sec:back} presents the background on GHCI and introduces the dimensions of the framework.
%Section \ref{sec:RQ} introduces the goals and motivations of the study. 
Section \ref{sec:schema} presents a translation of the four dimensions in the GHCI to SE human capital.
The systematic mapping methodology is then presented in Section \ref{sec:reviewMethod}, with the results to the research questions are presented in Section \ref{sec:results}.
Section \ref{sec:discussion} presents our indicators for the SE human capital.
%, which includes a discussion of the strengths and weaknesses of the study. 
Finally, we conclude the study in Section \ref{sec:conclusion}.

%\section{Background}
%\label{sec:back}
%The OSS Human Capital proposed in this dissertation is aimed at providing a synthetic assessment of a OSS's Human Capital.

\section{A Reconceptualization of Human Aspects in SE}
\label{sec:schema}
Inspired by the GHCI classification schema, Table \ref{tab:dimension-list} presents a summarized reconceptualization of the four dimensions into a SE context.
%Using the GHCI as a schema,
We now describe the rationale for each translated dimension from GHCI into SE human capital.

\begin{table}
\centering
\caption{Our proposed reconceptualization of human aspects classified into four human capital dimensions}
\label{tab:dimension-list}
\begin{tabular}{lp{44mm}p{44mm}}
\hline\noalign{\smallskip}
Dimensions & GHCI \citep{ghci2017} & SE Human Capital (proposed) \\
\noalign{\smallskip}\hline\noalign{\smallskip}
Capacity & Formal educational attainment as a result of past education investment. & \underline{Skill} attainment as a result of the past \underline{experiences} in software ecosystems.\\
\\
Deployment & Active participation in the workforce.
& Active participation in the \underline{community}. \\
\\
Development & Formal education of the next-generation workforce and continued upskilling and reskilling of the current workforce.
& Potential contributor involvement and upskilling and reskilling of \underline{contributors} in the \underline{community}.\\
\\
Know-how & Breadth and depth of specialized skills used in the workforce.
& Breadth and depth of specialized skills (i.e., \underline{work practices}) used in the \underline{community}. \\ 
\noalign{\smallskip}\hline
\end{tabular}
\end{table}

%the existing stock of education across generations, the Deployment subindex covers active participation in the workforce across generations, the Development subindex reflects current efforts to educate, skill and upskill the student body and the working age population, and the Know-how subindex captures the growth or depreciation of working-age people’s skillsets through opportunities for higher value-add work.

\paragraph{Capacity:}
A more educated population is better prepared to adapt to new technologies, innovate and compete on a global level.
Thus, GHCI defines capacity as formal education such as primary, secondary and tertiary levels of attainment \citep{ghci2017}.
%Education includes spontaneous action \cite{liu2016}. 
%Almost all of OSS development field excludes issues of education and relies on the spontaneous learning of contributors.
For open source software development, there is no formal education in general.
Instead, skills from experience is key.
A previous replication study reported that having a computer science or a software engineering background is not helpful during a requirement inspection, but having high experiences is significantly more effective \citep{SMS74}.
Considering these differences, we define capacity in SE human capital as the skill attainment of individuals from the past development experiences.

\paragraph{Deployment:}
Beyond formal learning, global human capital is enhanced in the workplace through learning-by-doing, tacit knowledge, exchange with colleagues and formal on-the-job learning. 
For GHCI, the deployment dimension measures how many people are able to participate actively in the workforce as well as how successfully particular segments of the population are able to contribute \citep{ghci2017}. 
The labor force participation of a country (i.e., employment rates) is the broadest measure of the share of its people participating in the labor market. 
In the SE context, abstraction is where the community involves developers that belong to a single project.
However, there are cases where the community covers a larger set of developers that contribute to an ecosystem of dependent projects. 
%This ecosystem may include the same programming language platform (i.e., npm developers) or have the same foundation (i.e., Apache Maven projects). 
We define the deployment dimension is understanding community structure or social interaction and measures how many contributors actively participate in the community.

\paragraph{Development:}
The development dimension in GHCI concerns current efforts to educate
%the formal education of
the next-generation workforce and continued upskilling and reskilling of the current workforce  \citep{ghci2017}.
Similarly, in our definition, the development dimension is
%knowing applications and accumulation of knowledge among the community of contributors, through understanding the growth of end-users or the learning curve of new contributors in the community.
potential contributor involvement, upskilling and reskilling of current contributors in the community.

\paragraph{Know-how:}
Know-how is concerned with the breadth and depth of specialized skills used in the workplace. 
In GHCI, the economic complexity is a measure of the degree of sophistication of a country's ``productive knowledge'' as can be empirically observed in the quality of its export products. 
In addition, the GHCI measures the current level availability of high- and mid-skilled opportunities and, in parallel, employer's perceptions of the ease or difficulty of filling vacancies \citep{ghci2017}.
In the SE context, we define the know-how dimensions as breadth and depth of specialized skill (i.e., work practices) used in the community, for example, the proportion of core or peripheral contributors' performance and knowledge loss caused by contributors' leaving the community.
%\noworries{please complete}
%Breadth and depth of specialized skills (i.e., work practices) use in the community.
%Maturity of work practice, core vs peripheral performance rates, knowledge loss, onbording
%\section{Research Methodology}
%%%%%%%%%%%%%%%%%%%%%%%%%%%%%%%%%%%%%%%%%%%%%%%
\begin{comment}

\end{comment}

\section{A Systematic Mapping of Human Capital in SE}
\label{sec:reviewMethod}

To realize SE human capital, we carried out a detailed systematic mapping to classify related studies into the four dimensions (i.e., presented in Section~\ref{sec:schema}).
A systematic mapping is more appropriate over a system literature review as (i) it is a repeatable method for identifying relevant studies to answer specific research questions \citep{Mendes-tse2010} and (ii) it is designed to give an overview of a research area through classification and counting contributions in relation to the categories of that classification \citep{Kitchenham2007, Petersen2008,Petersen-ist2015, Kitchenham-ease2010} and (iii) it does not have strict rules compared to systematic literature reviews; therefore, various types of papers can be covered.

%%%%%%%%%%%%%%%%%%%%%%%%%%%%%%%%%%%%%%%%%%%%%%%
\subsection{Research Questions}
\label{sec:RQ}
%%%%%%%%%%%%%%%%%%%%%%%%%%%%%%%%%%%%%%%%%%%%%%%
As part of the mapping study, we use the following research questions to describe and validate our reconceptualization of SE human capital:
\begin{description}
\item[$RQ_1$:]\rqtwo
Our intention is to investigate what human aspect topics and common SE terminology are often used to describe SE human capital.
\item[$RQ_2$:]\rqthree
We conjecture that there are different kinds of theories being analyzed and adapted for human aspect studies.
%Such evidence indicates areas where researchers have proposed interrelated concepts, definitions, and propositions to explain or predict events or situations related to specific dimensions of human capital.
Those theories indicate previous study interests in various human aspects from interrelated concepts, definitions, and propositions to explain or predict events or situations related to specific dimensions of human capital.
\item[$RQ_3$:]\rqfour
We would like to understand what kind of different data sources %of tangible assets (i.e., structural capital)
are used %by researchers
to uncover evidence of human capital. 
In detail, we extract data characteristics, such as data sizes, durations and diversities related to each dimension.
\end{description}
%%%%%%%%%%%%%%%%%%%%%%%%%%%%%%%%%%%%%%%%%%%%%%%

\subsection{Systematic Mapping Overview}
%In this section, we explain the different steps executed in our mapping study. 
%We first introduce our method for the systematic mapping study, which is based on the following characteristics \cite{Kitchenham2007}.
Similar to the systematic mapping study performed by \cite{SMS68}, %our method for the systematic mapping study strictly 
we considers the following characteristics recommended by \cite{Kitchenham2007}:
\begin{itemize}
\item ($C_1$) a defined search strategy
\item ($C_2$) a defined search string, based on a list of synonyms combined by ANDs and ORs
\item ($C_3$) a broad collection of search sources
\item ($C_4$) a strict documentation of the search
\item ($C_5$) quantitative and qualitative papers should be analyzed separately
\item ($C_6$) explicit inclusion and exclusion criteria
\item ($C_7$) paper selection should be checked by two researchers
\end{itemize}

%%%%%%%%%%%%%%%%%%%%%%%%%%%%%%%%%%%%%%%%%%%%%%%    
  	\begin{figure}[tb]
		 %\center
		 \small
			%\begin{minipage}{\linewidth}
			 \centering
			 \includegraphics[width=\linewidth]{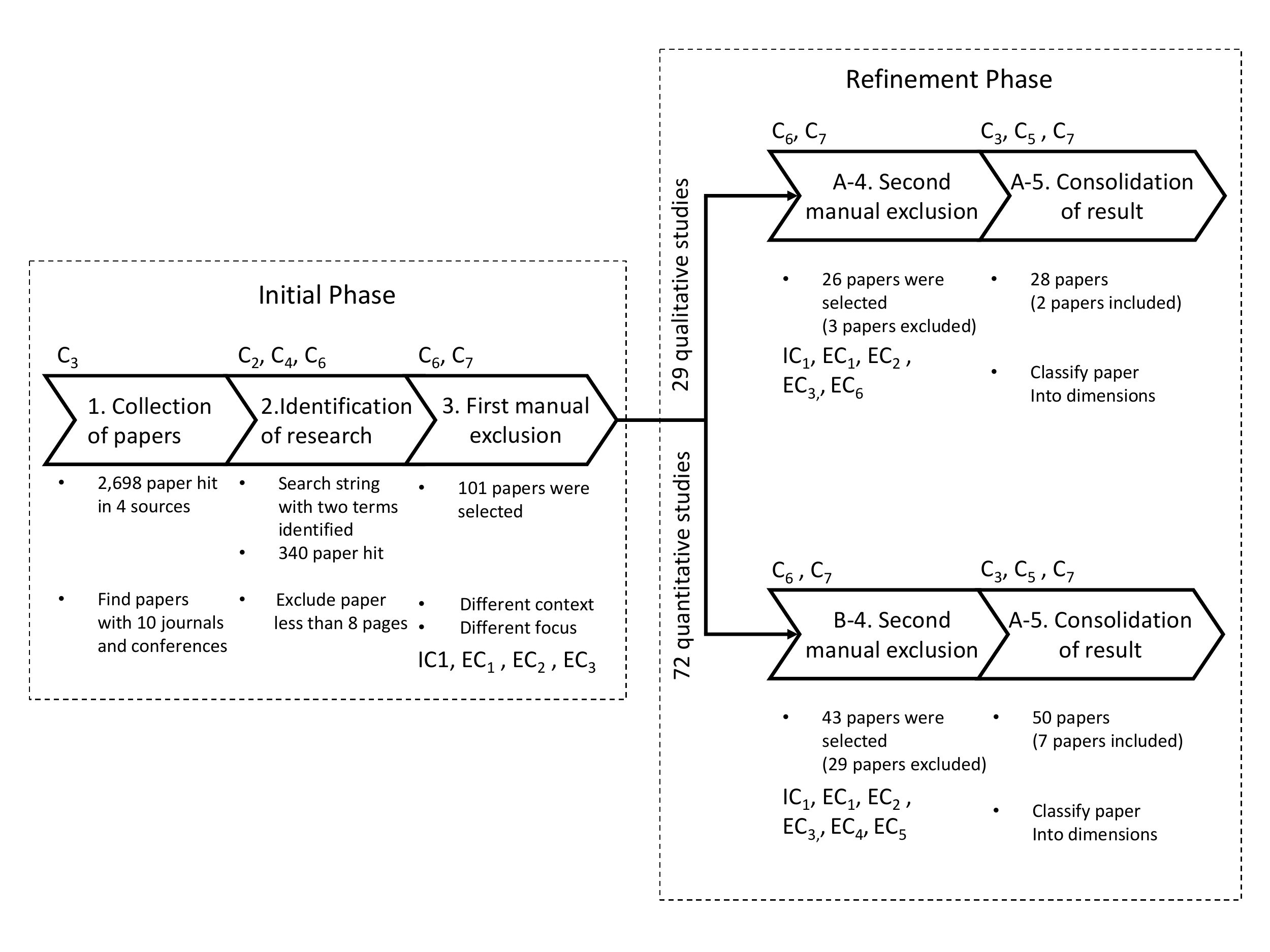}\\
			%\end{minipage}
		 \caption{Systematic Mapping Design}
		 \label{fig:mappingoverview}
	\end{figure}
%%%%%%%%%%%%%%%%%%%%%%%%%%%%%%%%%%%%%%%%%%%%%%%

Figure \ref{fig:mappingoverview} presents an overview of the mapping study design, which follows ($C_1$) a defined search strategy.
Our method is comprised of two parts, the initial phase and the refinement phase. 
We follow a total of five steps in our method.
Note that Steps 4 and 5 follow ($C_5$) two branches, dividing the papers into qualitative (A-4 and A-5) and quantitative studies (B-4 and B-5).  
We consider quantitative as studies include objective measurements and the statistical, mathematical, or numerical analysis of data collected, while the qualitative studies uncover trends in thought and opinions, and deeper analysis of the problem.
Qualitative studies were identified as having methods that use either a unstructured or semi-structured technique with focus on groups (group discussions), individual interviews, observations, and meta analysis. If a study consists of both quantitative and qualitative methods, we classify it as quantitative.
We now describe each step in detail.
%\myworries{Explanations needed. Related to ($C_5$). How about say quantitative and qualitative instead of method/survey?}

%\subsection{Initial Phase}
\subsubsection{Step 1: Collection of Papers}

%
%%%%%%%%%%%%%%%%%%%%%%%%%%%%%%%%%%%%%%%%%%%%%%%
\begin{table}
%\centering
\caption{Targeted SE journals and conferences with rankings and impact factors (IF) as of 2017}
\label{tab:pub-list}
%\scalebox{0.79}[0.79]{
\begin{tabular}{lll}
\hline\noalign{\smallskip}

Journal & (TSE) IEEE Transaction on Software Engineering  & IF: 2.63\\
                  & (EMSE) Empirical Software Engineering & IF: 3.28\\
                  & (ASEJ) Automated Software Engineering Journal & IF: 3.27\\
                  & (TOSEM) ACM Transactions on Software Engineering & IF: 2.87\\
                  & \hspace{10mm} and Methodology \\
                  & (IST) Information and Software Technology  & IF: 2.69\\
Conference & (ICSE) International Conference On Software Engineering & Rank: A*\\
                  & (ESEC/FSE) ACM Join European Software Engineering & Rank: A* \\
                  & \hspace{10mm} Conference and Symposium on Foundation of \\
                  & \hspace{10mm} Software Engineering \\
                  %& %(FSE) International Symposium on Foundations of Software Engineering \\
                  & (ICSME) International Conference on Software Maintenance & Rank: A \\
				  & (MSR) Working Conference on Mining Software Repositories & Rank: A\\
                  & \hspace{10mm} and Evolution \\
                  \noalign{\smallskip}\hline
%\multicolumn{3}{l}{http://www.core.edu.au}
\end{tabular}
%}
\end{table}
%%%%%%%%%%%%%%%%%%%%%%%%%%%%%%%%%%%%%%%%%%%%%%%
%
%%%%changed after draft*%%%%
Table \ref{tab:pub-list} shows the sources of papers for our mapping, along with their impact factors (IF) and conference rankings (the CORE Conference Ranking\footnote{http://www.core.edu.au} of 2017).
%%%%changed after draft*%%%%
%Table \ref{tab:pub-list} shows all 10 search sources for our mapping study.
To ensure a high quality of papers and to understand the state--of--the--art in the field, we specifically searched for papers in the top journals and conferences from the software engineering domain. 
To reduce its selection bias, we selected from a range of digital resources to follow ($C_3$) a broad collection of search sources: ACM Digital Library\footnote{\url{https://dl.acm.org/}}, IEEE Xplore\footnote{\url{http://ieeexplore.ieee.org/Xplore/home.jsp}},  Science Direct\footnote{\url{http://www.sciencedirect.com/}}, and Springerlink\footnote{\url{https://link.springer.com/}}.

As shown in Figure \ref{fig:mappingoverview}, we extracted 2,698 papers from the four search sources, which map to the top ten publication venues for software engineering.
Additionally, we only included technical papers, hence filtering out short papers, editorials, tutorials, panels, poster sessions and prefaces and opinions (i.e., we automatically filtered out any papers that was shorter than 8 pages).
Since our intention is to understand the current trends of human-related research, we collected papers that were published in the last five years (i.e., 2013 -- 2017).

\subsubsection{Step 2: Identification of Research}
To provide a comprehensive picture of recent research related to human capital, we used ($C_2$) a defined search string to identify the research area. In this step, we conducted the first automated ($C_6$) explicit inclusion and exclusion criteria based on the search results.

%%%%%%%%%%%%%%%%%%%%%%%%%%%%%%%%%%%%%%%%%%%%%%%    
  	\begin{figure}[tb]
		 %\center
		 \small
			%\begin{minipage}{1\linewidth}
			 \centering
			 \includegraphics[width=0.8\linewidth]{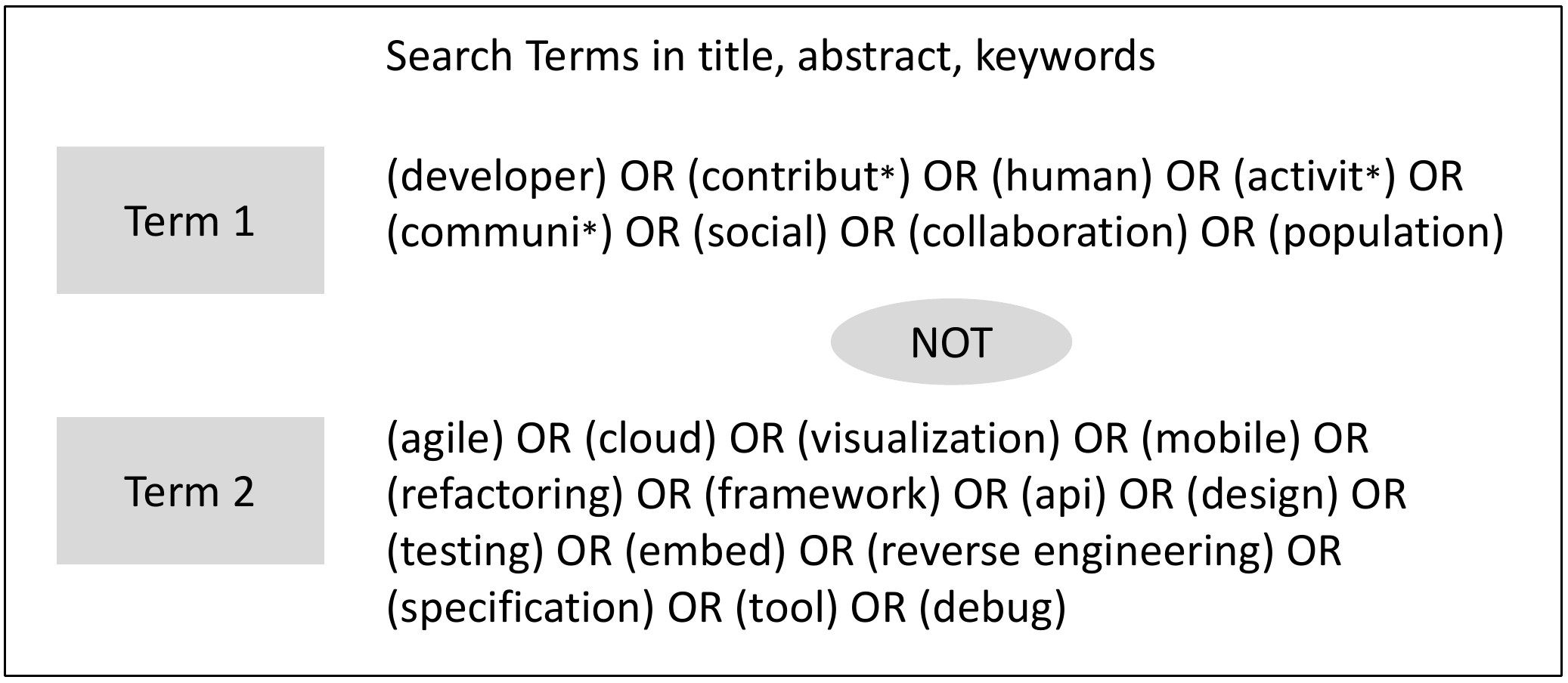}\\
			%\end{minipage}
		 \caption{Defined search strings}
		 \label{fig:term}
%Term 1: (developer) OR (contribut*) OR (human) OR (activit*) OR (communi*) OR (social) OR (collaboration) OR (population)
%Term 2:(agile) OR (cloud) OR (visualization) OR (mobile) OR (refactoring) OR (framework) OR (api) OR (design) OR (testing) OR (embed) OR (reverse engineering) OR (specification) OR (tool) OR (debug)
	\end{figure}
%%%%%%%%%%%%%%%%%%%%%%%%%%%%%%%%%%%%%%%%%%%%%%%

\begin{comment}
agile 			50
cloud			34
visualization	24
mobile			90
refactoring		62
middleware		17
framework		155
api				98
design			301
embed			33
reverse eng.	13
specification	94
tool			412
testing			210
debug			131
\end{comment}

Figure \ref{fig:term} shows the two terms (\texttt{term 1} and \texttt{term 2}) used in our search string.
To understand human-related areas in SE, we formulate \texttt{term 1} to include \textit{developer},  \textit{human}, \textit{social}, \textit{collaboration}, \textit{population}, and so on.
To capture synonyms and other extensions, as shown in Table \ref{tab:word}, we stem and use the keywords \textit{contribu*}, \textit{activit*} and \textit{communi*} to expand the search space.
Since, we are interesting in more generic human-related research from the SE domain, our search string also includes a exclusion list (i.e., \texttt{term 2}).
We exclude words in term 2 to avoid papers that are specific to particular research topics in SE, such as \textit{agile}, \textit{cloud}, \textit{reverse engineering}, and so on.
We apply these search strings to the title, abstract and keywords sections of targeted papers.
 
\begin{table}[tb]
\centering
\caption{Synonyms of Keywords used in the search.}
\label{tab:word}
\begin{tabular}{ll}
\hline\noalign{\smallskip}
Base Term & Synonyms \\
\noalign{\smallskip}\hline\noalign{\smallskip}
contribut$^{*}$ & contributor, contribution \\
activit$^{*}$ & activity, activities \\
communi$^{*}$ & community, communities, communication(s) \\
\noalign{\smallskip}\hline
\end{tabular}
\end{table}

%%%%%%%%%%%%%%%%%%%%%%%%%%%%%%%%%%%%%%%%%%%%%%%    
  	\begin{figure}[tb]
		 %\center
		 \small
			\begin{minipage}{1\linewidth}
			 \centering
			 \includegraphics[width=.6\linewidth]{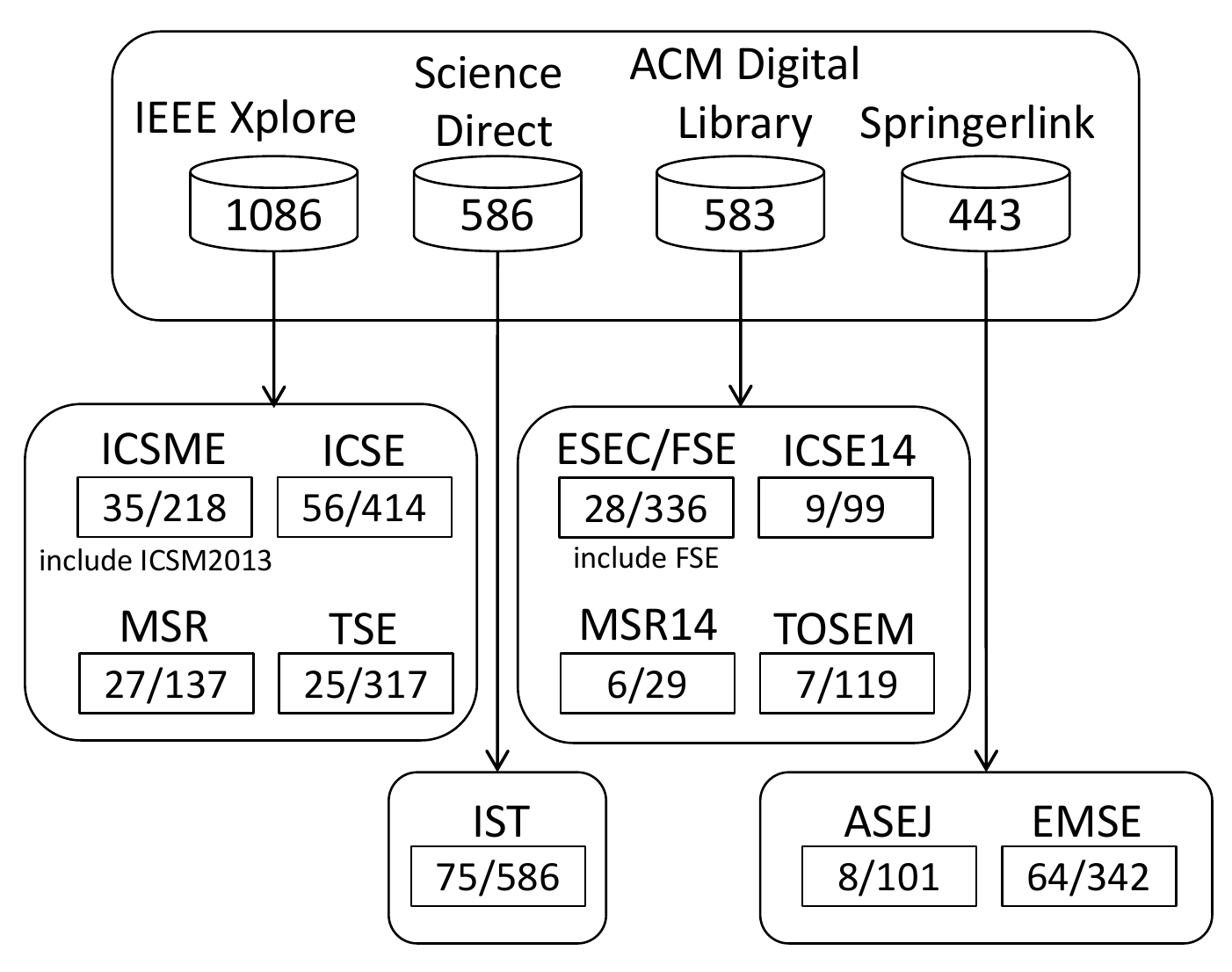}\\
			\end{minipage}
		 \caption{The process flow of the strict documentation procedure followed ($C_4$) to obtain 340 papers.}
		 \label{fig:collection}
	\end{figure}
%%%%%%%%%%%%%%%%%%%%%%%%%%%%%%%%%%%%%%%%%%%%%%%

As shown in Figure \ref{fig:mappingoverview}, we end with 340 papers after our automatic search execution.
For ICSE and MSR conferences, we found that special editions of ICSE14 and MSR14 were only published in the ACM Digital Library and not the Xplore.
Figure \ref{fig:collection} shows the details of remaining papers from the original 2,698 papers collected in Step 1.

\begin{comment}
Inclusion criteria
Articles that focus on software developers or contributors.

Exclusion criteria
Data: Analysis subjects are not contributors’ activity data.	
Evaluation: Purpose of objective experiment is not evaluation of a proposed method.
Purpose: Purpose of objective experiment is not to collect activity data.

Product-focused
Human recommendation
\end{comment}

\subsubsection{Step 3: First Manual Exclusion}
To complete the initial phase, Step 3 involves the manual exclusions from the collected 340 papers. 
This step involves ($C_6$) an explicit inclusion and exclusion criteria to remove papers that have a different context and focus to our research area.

For this manual exclusion, the following inclusion and exclusion criteria were applied to the abstract of each paper.

%\paragraph{Inclusion criteria}
\noindent
\textbf{Inclusion criteria}: 
Only a single inclusion criterion is defined, namely, ($IC_1$), the paper should focus on humans (i.e., software developers and contributors in a project).

%\paragraph{Exclusion criteria}
\noindent
\textbf{Exclusion criteria}:
Three exclusion criteria were defined that cover the datasets, purposes, and the evaluation of the studies. The following papers were excluded.
\begin{itemize}
\item ($EC_1$) the paper does not analyze any activity data
\item ($EC_2$) the purpose of the paper is collecting activity data
\item ($EC_3$) the paper intends to evaluate their proposed methods
\end{itemize}
Papers of survey and meta studies (in the qualitative category) were excluded if their primary studies meet the above exclusion criteria.
To reduce bias and follow ($C_7$), this manual paper selection was performed by the first and the second authors.
As a result of Step 3, we were able to reduce the collected 340 papers to 101 papers.

%\subsection{Refinement Phase}
%The refinement phase includes a separation between ($C_5$) quantitative and qualitative papers. 
%As a result, we divide our two steps into two branches (A and B).

%\subsection{Step 4: Second Manual Exclusion of Quantitative (i.e., method) and Qualitative (i.e., survey) papers}
\subsubsection{Step 4: Second Manual Exclusion of Qualitative and Quantitative papers}
Similar to Step 3, we use the same inclusion and exclusion criteria (i.e., $IC_1$, $EC_1$, $EC_2$ and $EC_3$) for the quantitative papers in Step B-4.
However, Step 4 includes a full reading of all contents of the paper.
Finally two exclusion criteria have been added to Step B-4:
\begin{itemize}
\item ($EC_4$) the paper focuses on software artifacts (i.e., products)
\item ($EC_5$) the paper proposes human recommendation techniques
\end{itemize}

For the qualitative papers in Step 4-A, we add another criterion for exclusion ($EC_6$) the paper focuses on specific techniques, systems, or phenomena which differs from the theme of human aspects.
In this manual analysis, we added the third author (i.e., making the total number of reviewers to three) to increase the confidence and the quality of our survey.
As a result of Step 4, we were able to reduce the initial 101 papers to 69 papers (26 qualitative and 43 quantitative papers).

\subsubsection{Step 5: Consolidation of Results}
In this step, three processes are performed. 
%As shown in Figure \ref{fig:mappingoverview}, after manual exclusion steps, 
First, we conduct a review of excluded papers, as recommended by \citep{Petersen-ist2015}.
Some papers (i.e., two qualitative and seven quantitative papers) were included that were excluded in the initial phase, bringing the final papers to 78 in total (29 qualitative and 49 quantitative papers).

%%%%%%%%%%%%%%%%%%%%%%%%%%%%%%%%%%%%%%%%%%%%%%%
  	\begin{figure}[tb]
		 \center
		 %\small
			%\begin{minipage}{\linewidth}
			 \centering
			 \includegraphics[width=0.8\linewidth]{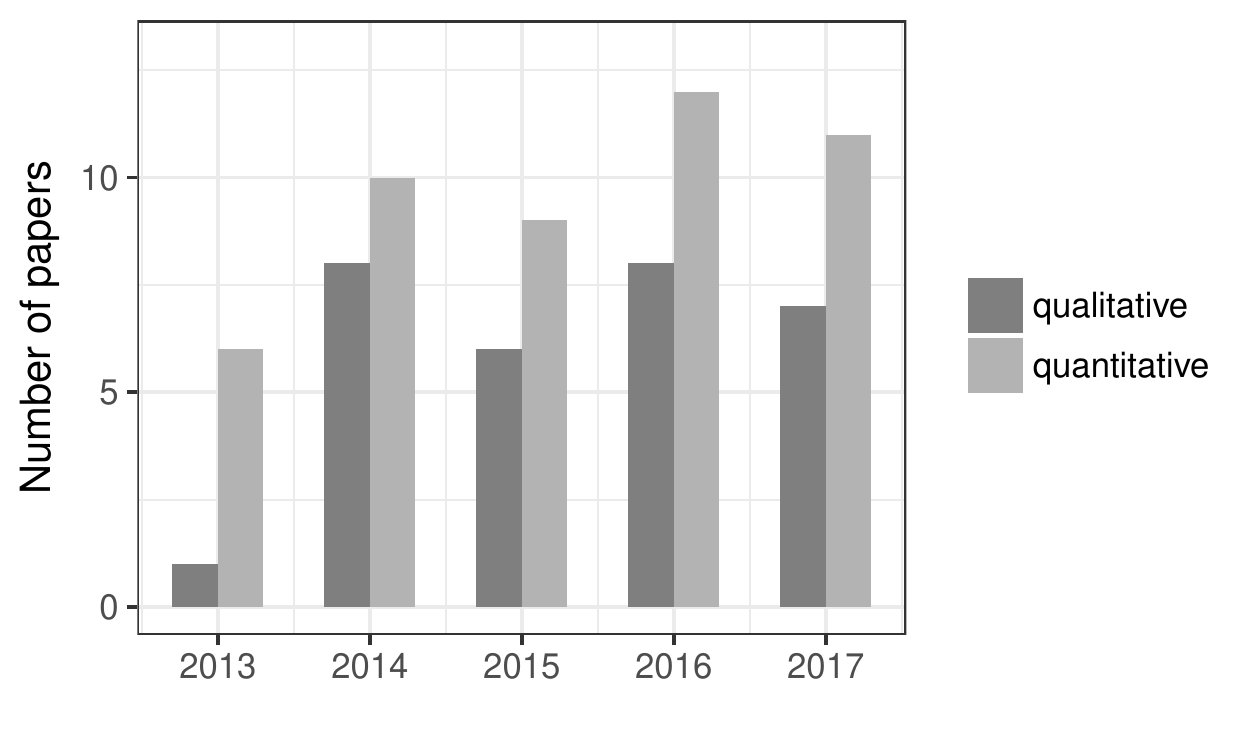}\\
			%\end{minipage}
		 \caption{The number of papers over the time period}
		 \label{fig:par-year}
	\end{figure}
%%%%%%%%%%%%%%%%%%%%%%%%%%%%%%%%%%%%%%%%%%%%%%%

%%%%%%%%%%%%%%%%%%%%%%%%%%%%%%%%%%%%%%%%%%%%%%%
  	\begin{figure}[tb]
		 \center
		 %\small
			%\begin{minipage}{\linewidth}
			 \centering
			 \includegraphics[width=0.8\linewidth]{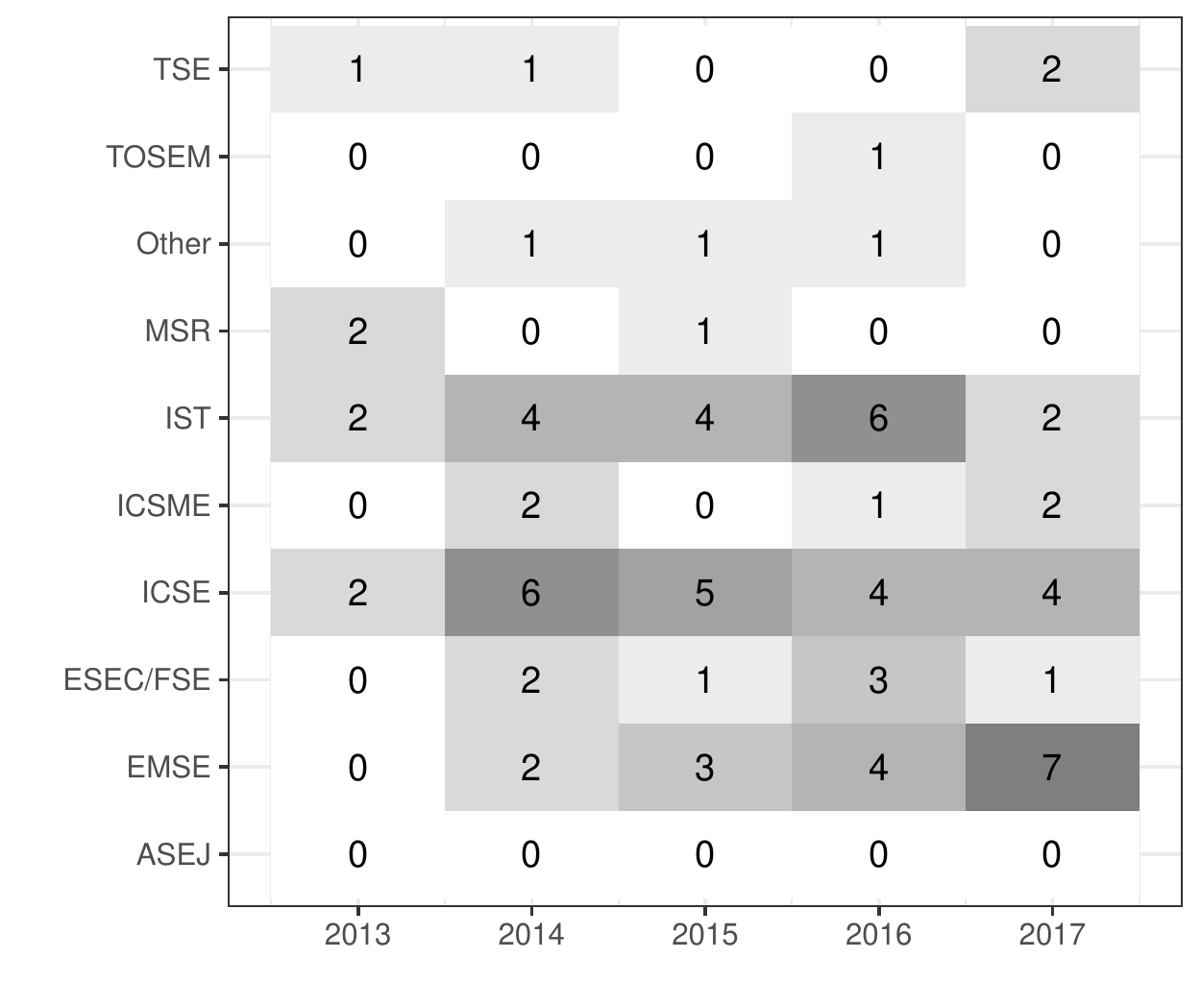}\\
			%\end{minipage}
		 \caption{The number of papers per conferences and journals from 2013 to 2017}
		 \label{fig:par-year2}
	\end{figure}
%%%%%%%%%%%%%%%%%%%%%%%%%%%%%%%%%%%%%%%%%%%%%%%

%Figures \ref{fig:par-year} and \ref{fig:par-year2} show the trends of human aspects.
Figure \ref{fig:par-year} shows the number of papers identified within the years 2013--2017 to each research category (qualitative and quantitative).
%We see that quantitative papers show a gradual increase year by year. 
We see that from 2014 both qualitative and quantitative papers have been consistently published.
%Conceivably survey paper in Human Capital area has some trend from one year to another.
Figure \ref{fig:par-year2} shows a heat map of the number of paper publications in each conference or journal in each year. 
This shows that many papers related to human aspects have been presented in IST, ICSE and EMSE.

%\begin{landscape}
\begin{table}[tb]
\centering
\caption{Summary of the paper classification into each dimension and quantitative and qualitative categories. Underlines indicate papers mutually inclusive in one or more dimensions.}
\label{tab:dimension}
\begin{tabular}{lll}
\hline\noalign{\smallskip}
& \multicolumn{2}{c}{Paper} \\
Dimension & \multicolumn{1}{c}{Quantitative} & \multicolumn{1}{c}{Qualitative} \\
\noalign{\smallskip}\hline\noalign{\smallskip}
%%%%%%%%%%%%%%%%%%%%%%%%%%%%%%%%%%%%%%%%%%
Capacity
& S09, S16, S18, S26, S39, \underline{S74} & S47, S65 \\
\noalign{\smallskip}
%%%%%%%%%%%%%%%%%%%%%%%%%%%%%%%%%%%%%%%%%%
Deployment
& S01, S02, S08, S12, S14, S17, & S07, S24, S28, S36, S38, S45, \\
& S19, S27, S29, S34, S42, S43, & S48, S59, S63 \\
& S44, \underline{S50}, S53, S66, S67, S69, & \\
& S70, S71, S76, S78 & \\
\noalign{\smallskip}
%%%%%%%%%%%%%%%%%%%%%%%%%%%%%%%%%%%%%%%%%%
Development
& S22, S31, S33, S49, S55, S72 & S03, S11, S46, S56, S57, S68 \\
\noalign{\smallskip}
%%%%%%%%%%%%%%%%%%%%%%%%%%%%%%%%%%%%%%%%%%
Know-how
& S05, S10, S15, S20, S21, S23, & S04, S06, S13, S25, S37, S41, \\
& S30, S32, S35, S40, \underline{S50}, S54,& S51, S52, S60, S61, S62, S77 \\
& S58, S64, S73, \underline{S74}, S75 \\ 
\noalign{\smallskip}\hline
\end{tabular}
\end{table}

Next, three reviewers classified all collected papers into the dimensions. 
A consensus was reached among all the reviewers on which dimension best described the paper. 
%From these papers, key indicators were then derived that are consistent during the classification process. 
Our classifications allow for mutually inclusive, thus papers can belong to multiple dimensions.
We used the following rationales used to classify each paper, which were based on the rationales presented in Section \ref{sec:schema}. 
\begin{itemize}
\item \textbf{Capacity} - Papers that discuss learning experiences of contributors.
\item \textbf{Deployment} - Papers that discuss contributor participation.
\item \textbf{Development} - Papers that discuss potential contributor involvement and learning within their community.
\item \textbf{Know-how} - Papers that discuss work practices in their community.
\end{itemize}
Table \ref{tab:dimension} shows the results of the classifications of the collected papers (the papers are presented in Appendix~\ref{sec:papers}).
Results indicate that much research has been carried out on the \textit{deployment} and \textit{know-how} dimensions.

%To address each of the research questions, the following approach was applied to analyze the consolidated papers. 
%Each research question identifies characteristics within the four dimensions, with the end goal a mapping from each viewpoint.
\textit{}\section{Results of the Mapping}
\label{sec:results}
%\subsection{Systematic Literature Review execution}
%%%%%%%%%%%%%%%%%%%%%%%%%%%%%%%%%%%%%%%%%%%%%%%
\begin{comment}
By executing the search on selected 10 journals and conferences, we collected a total of 2783 papers in our primary studies data. 
We applied including/excluding keywords, we retrieved a total of 340 papers in our primary studies.
The papers that were totally irrelevant were filtered after step ? of study selection process. We were then left with 101 relevant papers. 
After the checks from step ? of inclusion/exclusion criteria 69 studies remained.
%After checking the excluded papers, 9 studies were added to our systematic mapping.
%a total of 78 studies were left from primary search results. 
%Out of 44 relevant papers, 5 were excluded based on their low quality when evaluated against our quality assessment checklist (Section 4.4 and Appendix B) and 2 were found to be plagiarizing the work of two other papers already included. 
%So we were finally left with 00 studies. %(Appendix A, S1\UTF{2013}S37). 

We then performed step ?\UTF{2013}? of secondary search strategy to ensure the completeness of our results (Section ?). 
We retrieved further 9 studies that were relevant and were missing in primary search results. 
%After this phase our total number of included papers raised to 78 studies. 
After this step we ended up with a total of 78 studies for our final inclusion. 
\end{comment}
In this section, we present and discuss results of the systematic mapping in terms of human aspect topics (i.e., $RQ_1$), theories (i.e., $RQ_2$) and data (i.e., $RQ_3$) that were classified into SE human capital.
For each research question, we first introduce the approach to answer before we present each result.

%%%%%%%%%%%%%%%%%%%%%%%%%%%%%%%%%%%%%%%%%%%%%%%
%   	\begin{figure}[tb]
% 		 %\center
% 		 \small
% 			\begin{minipage}{1.0\linewidth}
% 			 \centering
% 			 \includegraphics[keepaspectratio,scale=0.5]{network.pdf}\\
% 			\end{minipage}
% 		 \caption{Overall Network of Topics used in our consolidated papers.}
% 		 \label{fig:network}
% 	\end{figure}
%%%%%%%%%%%%%%%%%%%%%%%%%%%%%%%%%%%%%%%%%%%%%%%

%%%%%%%%%%%%%%%%%%%%%%%%%%%%%RQ1%%%%%%%%%%%%%%%%%%%%%%%%%%%%
\subsection{$RQ_1$: Topics Discussed in SE Human Capital}

%%%%%%%%%%%%%%%%%%%%%%%%%%%%%%%%%%%%%%%%%%%%%%%
\begin{figure}[t]
 \subfloat[capacity]{\includegraphics[width=.4\linewidth]{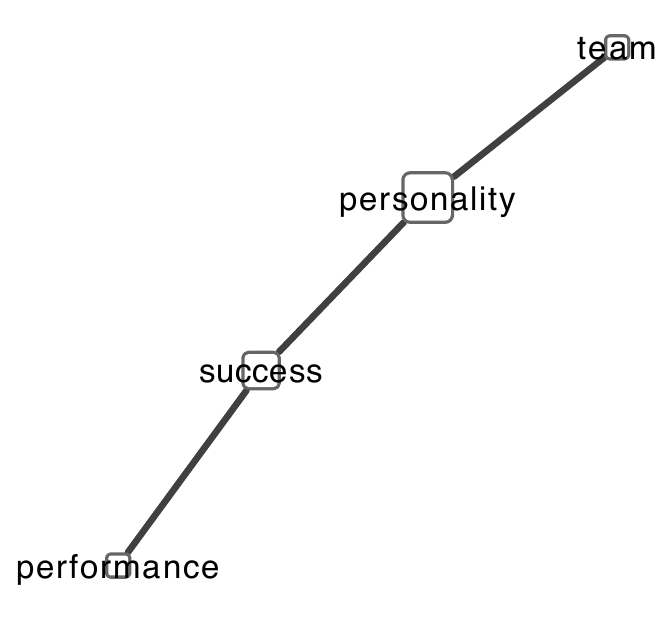}}  
 \subfloat[deployment]{\includegraphics[width=.6\linewidth]{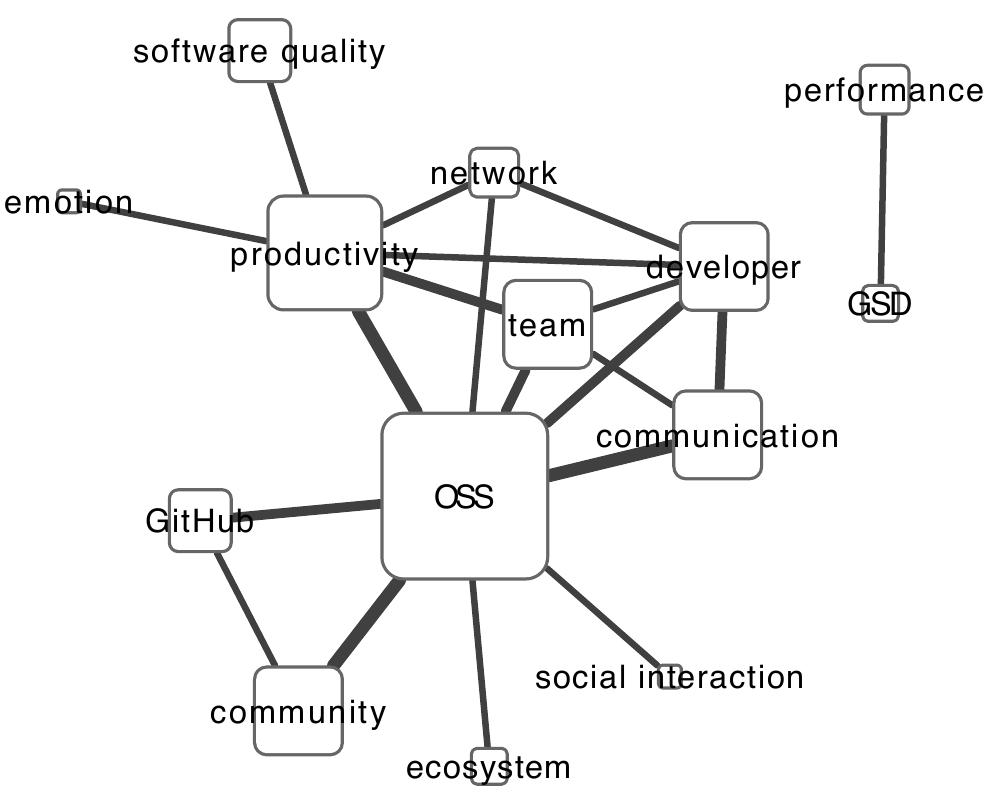}}
 \\  
 \subfloat[developent]{\includegraphics[width=.5\linewidth]{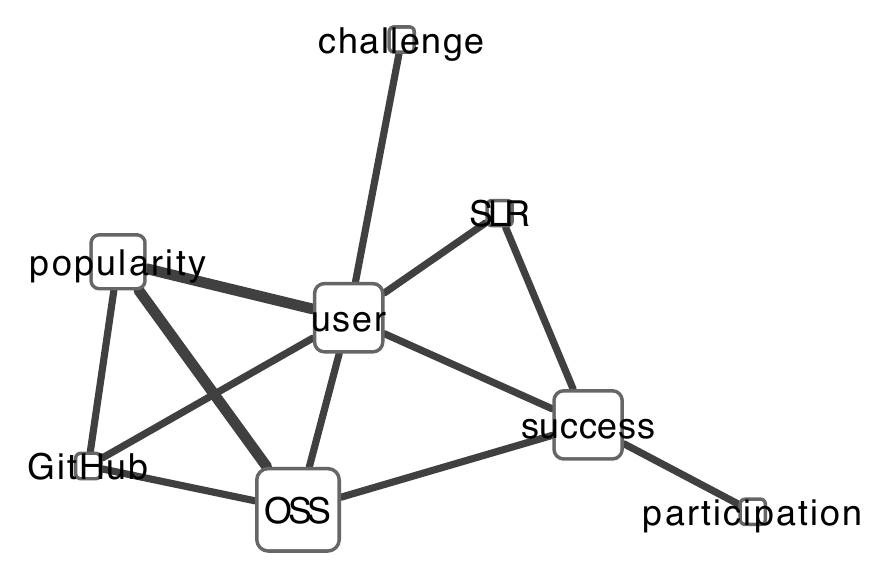}} 
 \subfloat[know-how]{\includegraphics[width=.5\linewidth]{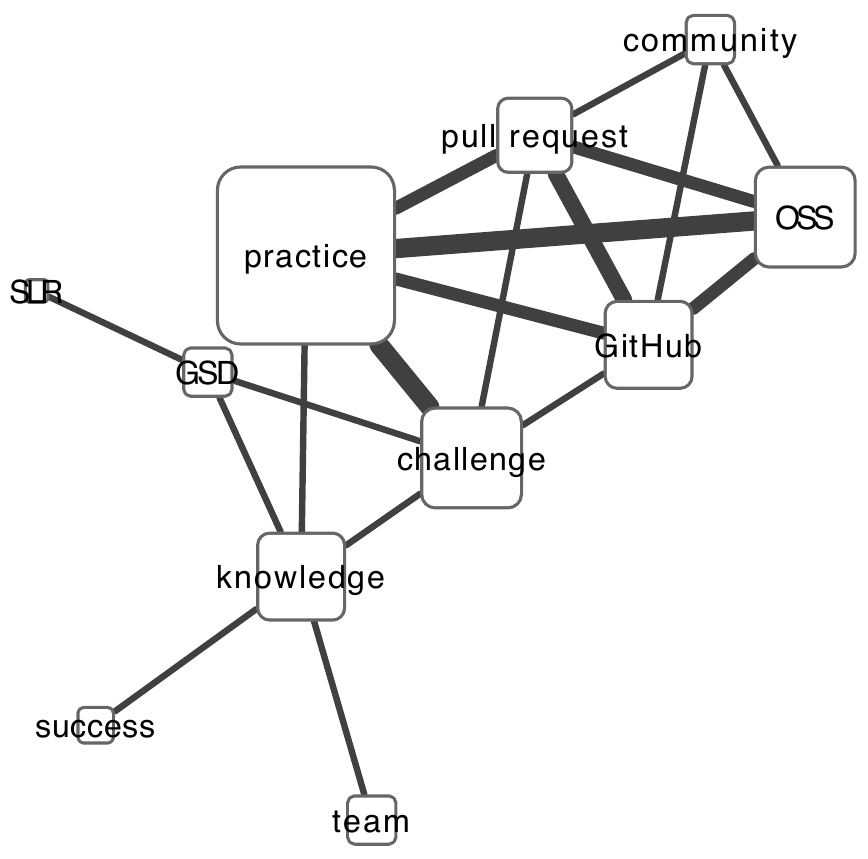}}  
    \caption{Topic co-occurrence networks for each dimension. The size of nodes represent the frequencies of topics, and the width of edges represent the frequencies of topics co-occurrences. GSD represents `global software development' and SLR is `systematic literature review'.}
	\label{fig:network_know-how}
\end{figure}
%%%%%%%%%%%%%%%%%%%%%%%%%%%%%%%%%%%%%%%%%%%%%%%

%%%%%%%%%%%%%%%%%%%%%%%%%%%%%%%%%%%%%%%%%%%%%%%
\begin{table}[t]
\centering
\caption{Topic pairs with high co-occurrence.}
\label{tab:topicsTop}
\begin{tabular}{llc}
\hline\noalign{\smallskip}
Dimension & Topic pair & \# co-occurrence \\
\noalign{\smallskip}\hline\noalign{\smallskip}
Capacity
& personality \& team & 2 \\
& personality \& success & 2 \\
& performance \& success & 2 \\
\noalign{\smallskip}
Deployment
& OSS \& community & 4 \\
& OSS \& communication & 4 \\
& OSS \& productivity & 4 \\
\noalign{\smallskip}
Development
& popularity \& user & 3 \\
& popularity \& OSS & 3 \\
\noalign{\smallskip}
Deployment
& practice \& challenge & 6 \\
& practice \& OSS & 6 \\
& pull request \& GitHub & 5 \\
\noalign{\smallskip}\hline
\end{tabular}
\end{table}
%%%%%%%%%%%%%%%%%%%%%%%%%%%%%%%%%%%%%%%%%%%%%%%

\textbf{Approach to answer $RQ_1$}.
Our approach includes a visualization and n-gram modeling to show what topics are discussed for the different dimensions.
\cite{Kuhrmann2017} reported the usefulness of \textit{word clouds} (tag clouds) for dataset cleaning in their systematic literature study experiences. 
Word clouds visualize the occurrences of words or terms in documents. Similar to their ideas of automatically extracting keywords as references,
we extract n-gram terms in a paper as topics by applying \texttt{n-gram IDF} \citep{Shirakawa:2015,Shirakawa:2017:IWN:3077622.3052775}. 
%Inverse Document Frequency (IDF) has been widely used in many applications because of its simplicity and robustness; however, IDF cannot handle phrases that are composed of more than one term. Because IDF gives more weight to terms occurring in fewer documents, rare phrases are assigned more weight than good phrases that would be useful in text classification.
%
N-gram IDF is a theoretical extension of Inverse Document Frequency (IDF) for handing multiple terms and phrases by bridging the theoretical gap between term weighting and multi-word expression extraction~\citep{Shirakawa:2015,Shirakawa:2017:IWN:3077622.3052775}.
%
%In software engineering research area, 
\cite{arm-icsme2017} reported that n-gram keywords detected with n-gram IDF were useful for bug report classification.
N-gram Weighting Scheme tool\footnote{https://github.com/iwnsew/ngweight} is used to extract n-gram topic keywords from title, abstract, and keywords in the collected papers.
Using a list of obtained n-gram keywords as a reference, we identified two to five keywords as topics for each paper. After integrating different expressions into the same terms (`open source software' is replaced with OSS, for example), co-occurrences of topics are analyzed. 

For the analysis results, we display topics within a dimension as connected networks, with the nodes representing the frequencies of topics and the width of the edges representing the frequency of topic co-occurrences. 
Sets of topics (keywords) for the collected papers are presented in Appendix~\ref{sec:papers}.

%Topics are extracted from the title, abstract, and keywords in the primary study and co-occurrences analyzed to answer the first research question.
%For easier interpretation of the results from our corpus of topics, an N-gram weighting scheme tool is used\footnote{https://github.com/iwnsew/ngweight}. 
%This tool uses enhanced suffix array \citep{Ibrahim-2004} to enumerate valid N-grams. 
%The output after applying the N-gram IDF tool to the pre-processed data is an N-gram dictionary, which is a list of all valid N-gram key terms. 
%Similar to Terdchanakul et al., \citep{arm-icsme2017}, we use the N-gram technique to formulate the most frequent keywords used in our corpus (i.e., title, abstract, keywords). 

%Results will be a generated network of topics, with the edges representing the co-occurrence score. 
%
\textbf{Results}.
Findings show that the extracted topic keywords correspond to the definitions of SE human capital (see Table \ref{tab:dimension-list}), thus providing a validation as well as insights into each dimension.
Figure~\ref{fig:network_know-how} presents topic co-occurrence networks for each dimension.
These networks provide insights into frequently studied topics.
We can see individual frequent keywords in each dimension and their relations in the networks.
Furthermore, Table~\ref{tab:topicsTop} summarizes frequently appeared topic keyword pairs in the same papers for the four human capital dimensions.
In \texttt{capacity}, individual \textit{personality} linking to \textit{team} or \textit{success}, or \textit{performance} and overall \textit{success} have been mainly studied.
In terms of \texttt{deployment}, \textit{community}, \textit{communication}, and \textit{productivity} in \textit{OSS} have been widely targeted for the studies.
For \texttt{development}, \textit{popularity} among \textit{user}s, and \textit{popularity} of \textit{OSS} projects have been major interests.
Finally, in \texttt{know-how},
%The size of the network is influenced by either the number of consolidated papers within the dimension and diversity of topics used in the field.
%Complementary, Table \ref{tab:topicsTop} shows the main topics (i.e., top 3) topics for the collected papers. 
%For capacity, the topics relate to \textit{personality, team, success and performance}. 
%This result indicates that capacity is a more a team-based factor that is linked to \textit{success} and \textit{performance}. 
%In terms of deployment, the networks show that the topics have complex interactions, indicating that these common topics are discussed across the collected papers.
%The four common topics discussed are \textit{OSS, community, communication and productivity}. 
much of the topics are related to \textit{practices} as seen in Figure~\ref{fig:network_know-how}.
The topics of \textit{practice}s in \textit{OSS} and their \textit{challenge}s have been mainly studied.
In addition, \textit{GitHub}-related research have been popular, with links to \textit{pull request}s.

%%%%%%%%%%%%%%%%%%%%%%%%%%%%%RQ2%%%%%%%%%%%%%%%%%%%%%%%%%%%%
\subsection{$RQ_2$: Theories Analyzed in SE Human Capital}
\begin{comment}
%%%%%%%%%%%%%%%%%%%%%%%%%%%%%%%%%%%%%%%%%%%%%%%
  	\begin{figure}[t]
		 %\center
		 \small
			\begin{minipage}{0.9\linewidth}
			 \centering
			 \includegraphics[keepaspectratio,scale=0.7]{theory.pdf}\\
			\end{minipage}
		 \caption{}
		 \label{fig:theory}
	\end{figure}
%%%%%%%%%%%%%%%%%%%%%%%%%%%%%%%%%%%%%%%%%%%%%%%
\end{comment}

%%%%%%%%%%%%%%%%%%%%%%%%%%%%%%%%%%%%%%%%%%%%%%%
\begin{comment}
\begin{table}[tb]
\centering
\caption{Coverage of selected papers with a theory}
\label{tab:notheory-proportion}
\begin{tabular}{lcl}
\hline \hline    
Dimension & Coverage of Papers & Paper id \\
\hline 
Capacity & 25\% &  \\
Deployment & 84\% &  \\
Development & 100\% &  \\
Know-how & 76\% &  \\
ALL &77\%\\
\hline
\end{tabular}
\end{table}
\end{comment}

\begin{table}[tb]
\centering
\caption{Percentages of the collected papers discussing theories}
\label{tab:theory-proportion}
\begin{tabular}{lrl}
\hline\noalign{\smallskip}
Dimension & Percentage & Paper \\
\noalign{\smallskip}\hline\noalign{\smallskip}
Capacity & 75\% (6/8) & S09, S16, S18, S26, S39, S65 \\
Deployment & 16\% (5/31) & S01, S14, S34, S67, S71 \\
Development & 0\% (0/12) &  \\
Know-how & 24\% (7/29) & S06, S23, S32, S40, S52, S54, S58 \\
\noalign{\smallskip}\hline
\end{tabular}
\end{table}
%%%%%%%%%%%%%%%%%%%%%%%%%%%%%%%%%%%%%%%%%%%%%%%

%%%%%%%%%%%%%%%%%%%%%%%%%%%%%%%%%%%%%%%%%%%%%%%
\begin{table}[tb]
\centering
\caption{Theories in the selected papers. Underlines indicate papers mutually inclusive, showing that multiple theories are employed.}
\label{tab:theory-paper}
\begin{tabular}{ll}
\hline\noalign{\smallskip}
Theory & Paper \\
\noalign{\smallskip}\hline\noalign{\smallskip}
Psychology/Psycholinguistics & S01, S06, S09, S16, S18, S39, S54, S65, S71 \\
Game theory  & S14, \underline{S67} \\
Group dynamics & \underline{S52}, \underline{S67}\\
Organization theory & \underline{S52}, S71\\
Demography & S34\\
Food web (ecology) & S23 \\
Financial risk management & S40 \\
Information field theory & S26  \\
Knowledge-based theory of the firm & S58 \\
Signaling theory & S32\\
\noalign{\smallskip}\hline
\end{tabular}
\end{table}
\textbf{Approach to answer $RQ_2$}.
Our approach includes a manual reading of all 78 collected papers, extracting each theory and grouping them into common types.
Similar to the mapping study, the first author manually extracts the type of theory used, looking for explicit keywords or a clear description of the theory in each paper.
Later, the results are validated with a consensus by the other co-authors.
Note that papers can contain multiple theories.
This is common, especially for empirical studies, that may employ mixed methods and various techniques and test several theories in a large study.
Analysis of the results includes a frequency count of each type of theory as well as their groupings into their respective dimensions.

\textbf{Results}.
Findings show that although a variety of theories are employed in SE research, many SE human aspect studies do not explicitly state their underlying theories.
Out of the 78 collected papers, only 18 papers described theories in their studies.
Grounded theory is excluded because it is a systematic methodology rather than a specific theory.
%This leaves 77\% of the papers not reporting any theory. 
We see in Table \ref{tab:theory-proportion} that most papers categorized in \textit{capacity} discussed theories, and
%most papers' theories were used for the capacity dimension.
%Interestingly, none of the papers classified from the development dimension reported a theory usage.
no paper in \textit{development} refers any theories.

%Complementary, 
Table \ref{tab:theory-paper} summarizes theories analyzed in the identified 18 papers.
We found 10 theories including game theory, organization theory, signaling theory, and so on.
%shows that most (i.e., 9) papers had adapted theory from the psychology/psycholinguistics field. 
Psychology/psycholinguistics is the most popular; it was referred in nine papers.
For instance, there were several papers that borrowed concepts from various aspects of the human psychology (i.e., S01 -- creation of models for team leader roles with personality types and gender classifications, S06 -- the success factors in global software development studied with a questionnaire, and S09 -- an experiment of personality factors and group processes).
Although some theories were adopted in multiple papers, six theories appeared only in single papers.

\begin{comment}
\myworries{Figure \ref{fig:theory-heat} shows the types of research methods used in complementary to each dimension. 
In this result, the experimental research method is the most common type of research method among the collected papers, and it is also used in the capacity papers. 
In addition, in action research, case study and surveys are popularly used with their theories. 
Action research represents research methods that research either initiated to solve an immediate problem or to improve the way issues are addressed and problems solved \cite{AR}.}
\end{comment}

%%%%%%%%%%%%%%%%%%%%%%%%%%%%%RQ3%%%%%%%%%%%%%%%%%%%%%%%%%%%%
\subsection{$RQ_3$: Data Sources Used in SE Human Capital}

\begin{table}[tb]
\centering
\caption{Summary of data sources of SE human capital into the four dimensions. Underlines indicate papers mutually inclusive, showing that multiple data sources where used in the papers. Note that VCS includes other version control system other than git. Other abbreviations are Bug Tracking System (BTS) and Issue Tracking System (ITS).}
%``other'' includes company data, stack overflow, code review repository, web document, gerrid and Ruby API.}
\label{tab:data_paper}
\begin{tabular}{lcll}
\hline\noalign{\smallskip}
Dimension & Percentage & Source & Paper \\%& Coverage \\
\noalign{\smallskip}\hline\noalign{\smallskip}
Capacity & 13\% (1/8) & Jazz  & S16 \\%& 100\%\\ 
\noalign{\smallskip}
Deployment & 68\% (21/31)
 & VCS & \underline{S12}, S27, \underline{S42}, \underline{S66}, \underline{S69}, S70, S76 \\%& 37\%\\ 
& & ML/Chat & \underline{S12}, S17, \underline{S19}, \underline{S29}, \underline{S42}, S64, \underline{S66}, S67, \underline{S69} \\%& 47\%\\ 
& & GitHub & \underline{S02}, S14, S34, S53, S71 \\%& 26\%\\
& & BTS/ITS & S08, \underline{S19}, \underline{S29}, \underline{S69} \\%& 21\% \\
& & Other & \underline{S02}, \underline{S19}, S43, \underline{S50}, S78 \\%& 21\%\\
\noalign{\smallskip}
Development & 50\% (6/12)
& VCS & S31, S49 \\%& 50\%\\
& & ML/Chat &  S33 \\%& 25\%\\ 
& & GitHub & S55 \\%& 25\%\\
& & BTS/ITS & \underline{S72} \\%& 25\%\\ 
& & Other & S22, \underline{S72} \\%& 50\%\\
\noalign{\smallskip}
Know-how & 55\% (16/29)
& VCS & \underline{S15}, \underline{S20}, \underline{S21}, \underline{S23}, S40, \underline{S64}, \underline{S75} \\%& 50\%\\ 
& & ML/Chat & \underline{S15}, \underline{S64}, \underline{S75} \\%& 25\%\\ 
& & GitHub & S05, \underline{S21}, S30, \underline{S32}, S35\\%& 42\%\\
& & BTS/ITS & \underline{S20}, \underline{S23}, \underline{S32}, \underline{S64}, \underline{S73} \\%& 42\%\\ 
& & Jazz & S54 \\%& 8\%\\
& & Other & S10, \underline{S50}, S58, \underline{S73}, \underline{S75} \\%& 25\%\\
\noalign{\smallskip}\hline
\end{tabular}
\end{table}
%%%%%%%%%%%%%%%%%%%%%%%%%%%%%%%%%%%%%%%%%%%%%%%

%%%%%%%%%%%%%%%%%%%%%%%%%%%%%%%%%%%%%%%%%%%%%%%
\begin{comment}
\begin{table}[tb]
\centering
\caption{Summary of Data Collection of Sources
}
\label{tab:dataDetails}
\begin{tabular}{llrrrrr}
\hline\noalign{\smallskip}
& & Capacity & Deployment & Development & Know-how & Percentage\\
\noalign{\smallskip}\hline\noalign{\smallskip}
Projects
& S ($>10$) & 1 & 9 & 3 & 8 & 58\%\\%21\\
& M ($11\sim 100$) & 0 & 7 & 0 & 1 & 22\%\\%8\\
& L ($101\sim$) & 0 & 3 & 1 & 3 & 19\%\\%7\\ 
\noalign{\smallskip}
%& Total & ? & ? & ? & ? & ?\\ \hline
%%%%%%%%%%%%%
Period
& S($<1 year$) & 0 & 1 & 2 & 3 & 16\%\\
& M($<5 year$) & 1 & 5 & 0 & 2 & 22\%\\
& L(5 $year\sim$) & 0 & 8 & 2 & 3 & 36\%\\
& **unspecified & 0 & 5 & 0 & 4 & 25\%\\
\noalign{\smallskip}
Origin
& Company & 1 & 2 & 1 & 2 & 16\%\\
& OSS & 0 & 17 & 3 & 10 & 83\% \\
\noalign{\smallskip}
Source
& Multi & 0 & 7 & 1 & 7 & 42\%\\
& Single & 1 & 12 & 3 & 5 & 58\%\\
\noalign{\smallskip}\hline
%\multirow{1}{*}{Total} & & 1 & 19 & 4 & 12 \\ \hline
\end{tabular}
\end{table}
\end{comment}
\begin{figure}[t]
\begin{center}
\includegraphics[width=.8\linewidth]{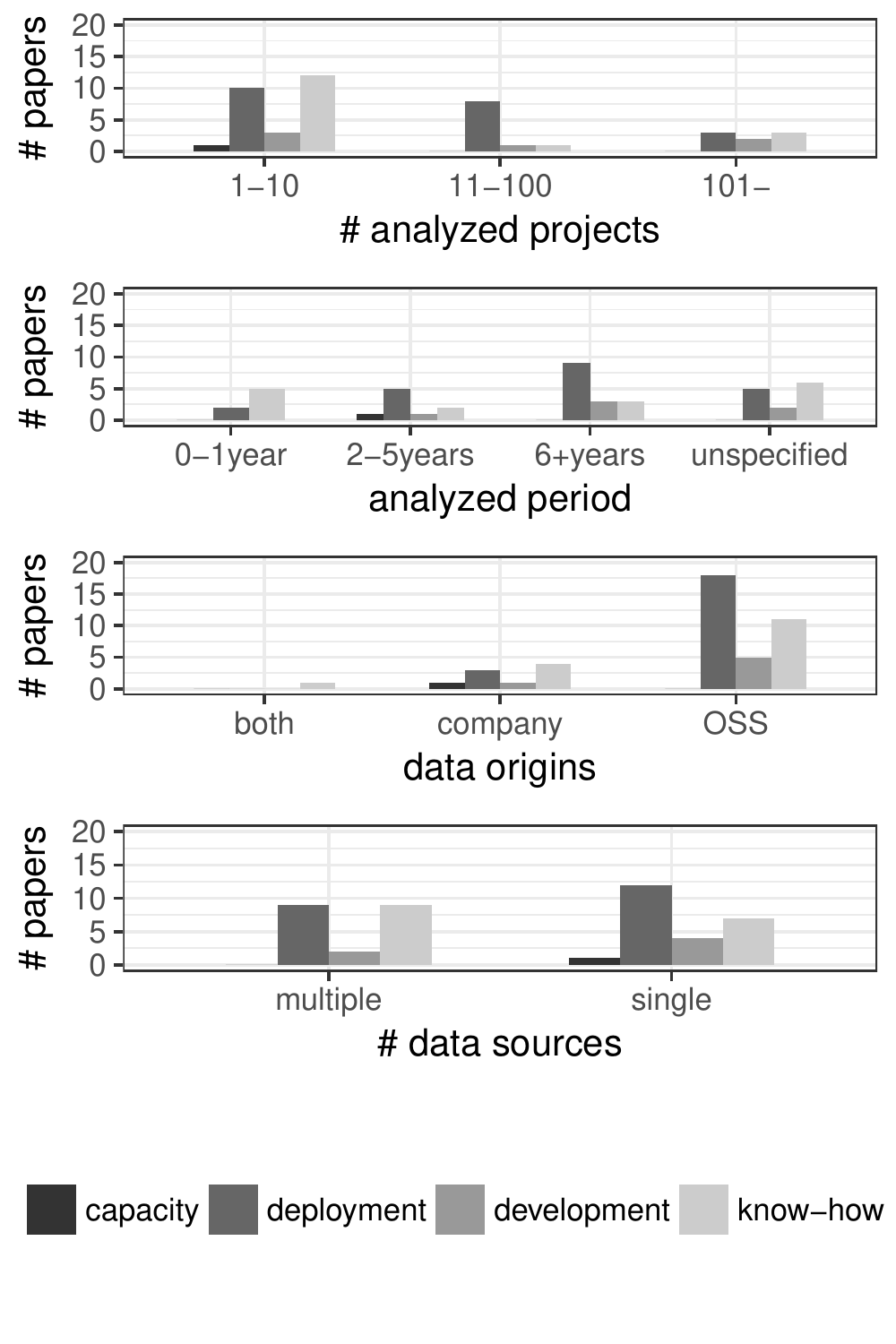}
\end{center}
% \subfloat[capacity]{\includegraphics[width=.5\linewidth]{proj.pdf}}  
% \subfloat[deployment]{\includegraphics[width=.5\linewidth]{period.pdf}}
% \\
% \subfloat[developent]{\includegraphics[width=.5\linewidth]{origin.pdf}} 
% \subfloat[know-how]{\includegraphics[width=.5\linewidth]{source.pdf}}  
    \caption{Sizes, periods, origins, and varieties of data sources of SE human capital in four dimensions.}
	\label{fig:dataDetails}
\end{figure}

%%%%%%%%%%%%%%%%%%%%%%%%%%%%%%%%%%%%%%%%%%%%%%%
\textbf{Approach to answer $RQ_3$}.
Similar to $RQ_2$, our approach includes a manual reading of the 49 quantitative collected papers, extracting each data source and grouping them.
Similarly, the first author manually extracts the data-source related information, then later reports to other co-authors for a validated consensus.
Firstly, we extract the type of data management system (source) from which the he data originated from (i.e., GitHub, version control system (VCS), Bug Traking System (BTS), Issue Tracking System (ITS) and so on.)
Note that papers can originate from multiple data sources.
Furthermore, we would like to understand other useful data information such as (a) the number of projects where analyzed in each paper, (b) the time period of the data collection, (c) whether the data is Open Source or company data and (d) whether the data is taken from multiple sources.
Analysis of the results includes a frequency count of each type of data source as well as their groupings into their respective dimensions. 

\textbf{Results}.
Findings show that although a variety of data sources are being employed in SE research, that analyze multiple project over a five to six year period.
Data sources are investigated from 49 quantitative papers. For reproducibility of measurement, we focus on archived data sources and exclude seven papers of experimentation studies (S01, S09, S18, S26, S39, S44, and S74).
Table \ref{tab:data_paper} summarizes data sources used in 42 quantitative papers.
%shows a breakdown of the data origins for the 36 consolidated papers (i.e., excluded survey papers). 
Note that \textit{others} refers to rarely used data sources in the collected papers, such as closed company data, stack overflow, code review, web documents, gerrit code review and Ruby API systems.
Compared to the percentages of theories used in Table~\ref{tab:theory-proportion}, the percentage of \textit{capacity} papers is the lowest, that is, most studies categorized in \textit{capacity} did not analyzed archived data sources. We consider that capacity-related studies using archived data is a challenging research topic and currently missing.
Papers belonging to the other three dimensions used various and common data sources, such as
version control systems (VCS), mailing lists (ML) and chat data, GitHub archives, bug tracking systems (BTS) and issue tracking systems (ITS).
%In this result, human capacity dimension papers have the least source of data (i.e., coverage of 3\%).
%From the table, it can be seen that researchers used assets such as mailing lists (ML), bug tracking (BTS) and issue tracking systems (ITS) to complement their code in their datasets (i.e., deployment = seven papers used code, while 17 papers used assets, development = two papers used code, while five papers used assets, know-how = six papers used code, while 12 papers used assets). 
%Additionally, most data originate from not only code but also from software assets. 

Figure \ref{fig:dataDetails} shows a detailed classifications of data sources (i.e., size as the number of analyzed projects, analyzed periods, data origins, and varieties of data sources) in 42 quantitative papers.
%Note that some papers were excluded from this analysis due to their difficulty of classification (i.e., S31, S33, S40, S75). 58\% of papers analyzed less than 10 projects in their study.
Although only a small number of projects analyzed single projects for their studies,
some papers used more than 100 projects.
%In fact, five of these papers only used a single project for analysis (i.e., S10 or S16). 
%On the other hand, one paper (i.e.,
For example, the paper S21 used 58,092 projects in their study.
Regarding analyzed periods, we found many papers analyzed more than five years data.
%Evidence then suggests that many of the papers collected data that ranged more than five years, with six papers having data ranging less than a year. 
Another interesting finding is that most papers used OSS projects (83\%) compared to closed company data.
In regards to data sources, about half of the papers used multiple sources, often combining code from other sources (i.e., as shown in Table \ref{tab:data_paper}) in their studies.

%\section{Discussion}
\section{Indicators for SE Human Capital}
\label{sec:discussion}
%In this section we discuss the implications of our study. 
%A key outcome of this study is a set of indicators for the different dimensions of human capital.
%Hence, we first introduce our set of indicators.
%Afterwards, in this section, we discuss the strengths and weaknesses of our study with threats to validity. 

\begin{table}[tb]
%\scalebox{0.9}[0.9]{
\centering
\caption{Proposed SE human capital indicators mapped to the collected papers. Underlines indicate papers mutually inclusive, showing that a  papers can contain multiple indicators.}
\label{tab:indicators}
\small
%\begin{tabular}{l|l|p{1.17cm}p{1.17cm}p{1.17cm}}
\begin{tabular}{llll}
\hline\noalign{\smallskip}
& & \multicolumn{2}{c}{Paper} \\
Dimension & Indicator & Quantitative & Qualitative \\
\noalign{\smallskip}\hline\noalign{\smallskip}
%%%%%%%%%%%%%%%%%%%%%%%%%%%%%%%%%%%%%%%%%%
%%%%%%%%%%%%%%%%%%%%%%%%%%%%%%%%%%%%%%%%%%%%%%%%%%%%%%%%%
Capacity
& Contributor activity profiling
& S09, S16, S18, & S47, S65 \\
& & S26, S39, \underline{S74} \\
\noalign{\smallskip}
%%%%%%%%%%%%%%%%%%%%%%%%%%%%%%%%%%%%%%%%%%
Deployment
& Community diversity
& S01 & S28, \underline{S38} \\
\noalign{\smallskip}
%%%%%%%%%%%%%%%%%%%%%%%%%%%%%%%%%%%%%%%%%%
& Community social interaction
& S02, \underline{S08}, \underline{S12}, & S45, S48, \underline{S63} \\ 
& & S17, \underline{S19}, S29, \\ 
& & \underline{S42}, \underline{S50}, \underline{S66}, \\
& & S67, S69, S70, \\
& & \underline{S71} \\
\noalign{\smallskip}
%%%%%%%%%%%%%%%%%%%%%
& Community structural complexity
& S14, \underline{S27}, S34, & S07, S36, S59 \\
\noalign{\smallskip}
%%%%%%%%%%%%%%%%%%%%%
& Participation rates
& S53, S78 & \underline{S38}, \underline{S63} \\ 
\noalign{\smallskip}
%%%%%%%%%%%%%%%%%%%%%
& Productivity rates
& \underline{S08}, \underline{S12}, \underline{S19}, & S24 \\ 
& & \underline{S27}, \underline{S42}, S43, \\ 
& & S44, \underline{S66}, \underline{S71} \\ 
\noalign{\smallskip}
%%%%%%%%%%%%%%%%%%%%%
%& Core vs. Peripheral Developer workload rates   
& Workload equality   
& S76 \\ 
\noalign{\smallskip}
%%%%%%%%%%%%%%%%%%%%%
Development
& Developer learning-curve
& S31, S49 & S11, S56 \\ 
\noalign{\smallskip}
%%%%%%%%%%%%%%%%%%%%%%%%%%%%%%%%%%%%%%%%%%
& End-user participation
& S22, S33, S55, & S03, S46, S57 \\
& & S72 & S68 \\
\noalign{\smallskip}
%%%%%%%%%%%%%%%%%%%%%
Know-how
& Core contributor knowledge
& S15, S20, S54 \\
\noalign{\smallskip}
%%%%%%%%%%%%%%%%%%%%%
& Knowledge loss rates
& S40 \\
\noalign{\smallskip}
%%%%%%%%%%%%%%%%%%%%%
& Maturity of work practices
& S05, S10, S23, & S04, S06, S13, \\ 
& & S30, S32, S35, & S25, S37, S41 \\ 
& & \underline{S50}, S58, S64, & S51, S52, S60, \\
& & S73, \underline{S74}, S75 & S61, S62, S77 \\
\noalign{\smallskip}
%%%%%%%%%%%%%%%%%%%%%
& Onboarding rates
& S21 \\
\noalign{\smallskip}\hline
\end{tabular}
%}
\end{table}
%%%%%%%%%%%%%%%%%%%%%%%%%%%%%%%%%%%%%%%%%%%%%%%%%%%%%%%%%

The key outcome of this study is a set of indicators for the different dimensions of SE human capital.
Based on the results of the mapping study and results of the three research questions, we were able to extract candidate indicators that map to each dimension.
%\subsection{Building Indicators for Human Capital}
Table \ref{tab:indicators} describes our proposed indicators and the mapping to their respective inspired collected papers.
This is another mapping of existing papers towards the future creation of a SE human capital index (SE-HCI); all 79 papers are categorized into 13 indicators.
We now describe the rationale and definition of each indicator by dimension and related studies.

%\vspace{-2mm}
\paragraph{Capacity Dimension}
\begin{enumerate}
%\vspace{-2mm}
\item \textbf{Contributer activity profiling} - This indicator captures the skill attainment based on individual contributor activities. 
One of related papers is a study about measuring team personality and climate (i.e., S09). 
This study measured the neuroticism, extroversion, conscientiousness and all that of developers from their experiment. 
Another study is related to personality profiles of developers (i.e., S16). 
This study analyzed message exchanges or developers' tasks using code and assets data source.
\end{enumerate}

%\vspace{-2mm}
\paragraph{Deployment Dimension}
\begin{enumerate}
%\vspace{-2mm}
\item \textbf{Community diversity} - This indicator measures the diversity within the community. 
For example, S01 studied team leadership roles with personality types and gender classification. 
It used experimental data to develop a model for software development team composition by keeping gender as a major effecting variable with personality. 
Furthermore, study S38 identified barriers for female participation on stack overflow by
interviewing female contributors about contribution barriers in online communities.

%\vspace{-2mm}
\item \textbf{Community social iteration} - The social interaction indicator is a measure of contributor collaboration and communication within the community.
Examples include a study about communication in open source software development mailing lists (i.e., S17) and a study about identification of contributors' collaborations from different sources by analyzing source code co-changes (i.e., S29).

\item \textbf{Community structural complexity} - This indicator focuses on the complexity of the community. 
An example is studying about population structure in OSS projects (i.e., S34).
This study created a population pyramid considering the contributors' activity periods in the community.
Another example is a study about the impacts of organizational factors on software quality (i.e., S59), where authors reported observations of an in-house software development project within a large telecommunications company.

%\vspace{-2mm}
\item \textbf{Participation rate} - This indicator describes contributor participation in a particular activity such as code review (i.e., S53, S78).
S53 investigated the phenomena of inactive code review contributors from activities such as pull requests, while S78 is a study about review participation in code review, introducing several metrics such as purpose, history, and prior activities of reviewers and patch authors.

%\vspace{-2mm}
\item \textbf{Productivity rate} - This indicator measures the productivity of contributors.
For instance, study S24 %evaluates developer performance for software-intensive products. This study
investigated performance measurement practices related to software product development activities by
interviewing managers how they perceive and evaluate performance in large organizations from a managerial perspective. 
On the other hand, study S44 is about sensing developers' emotions, progress and the use of biometric measures.

%\vspace{-2mm}
%\item \textbf{Core Developer vs. Peripheral Developer workload equality} - These rates are in regard to the work activities of developers in the community. 
\item \textbf{Workload equality} - This indicator focuses on workload of contributors in the community.
Example studies include observation of the variation and specialization of workload in an ecosystem community to identify developers' activity types and comparing the number of files that developers modify (i.e., S76).
\end{enumerate}

%\vspace{-2mm}
\paragraph{Development Dimension}
\begin{enumerate}
%\vspace{-2mm}
\item \textbf{Developer learning-curve} - This indicator is a developer-centric measurement of skill development.
For example, study S31 investigated the effect of the Google summer of code
by comparing developers' activities in OSS projects before and after participating the Google summer of code event.
Another study conducted a questionnaire to investigate the impressions, motivations, and barriers of one time code contributors to FLOSS projects (i.e., S56).

%\vspace{-2mm}
\item \textbf{End-user participation} - This indicator is a user-centric measurement of skill development.
For instance, previous study analyzed how end-users and their communities use public repositories (i.e., S22).
%This study analyzes end-user programmer communities, the characteristics of artifacts in community repositories, and how authors evolve over time.
Another study investigated factors that impact the popularity of GitHub repositories (i.e., S55).
This work analyzed stars awarded to GitHub projects and analyzed the popularity growth of these repositories. 
\end{enumerate}

%\vspace{-2mm}
\paragraph{Know-how Dimension}
\begin{enumerate}
%\vspace{-2mm}
\item \textbf{Core contributor knowledge} - This indicator explores the knowledge of core contributors (compared to peripheral contributors).
For instance, study S15 classified developers into core and peripheral by proposing network metrics.
%This study measures metrics related to commits and emails within the project.
Another example is a study about determining developers' expertise and roles (i.e., S20).
This study analyzed bug tracker and source code repositories to characterize developers.

%\vspace{-2mm}
\item \textbf{Knowledge loss rates} - This indicator investigates the loss of knowledge because of contributors' leaving from the community.
For instance, study S40 quantified and investigated how to mitigate turnover-induced knowledge loss.
In detail, it quantified the extent of abandoned source files using source code history and assessed knowledge loss.

%\vspace{-1mm}
\item \textbf{Maturity of work practices} - This indicator measures the degree of work practices used in contributors' software development processes. 
For instance, study S05 is concerned with a pull-based software development model. 
It explored how pull-based software development worked by analyzing pull requests and comments history.
On the other hand, study S10 explored the prior beliefs of developers at Microsoft, confirming beliefs to actual empirical data.
%It is a survey to understand a priori opinions on issues such as cost, quality, and interval related to the project.

%\vspace{-2mm}
\item \textbf{Onboarding rates} - This indicator measures the retention of contributors in the community.
%In this thesis, we identified a study that explores a precursor to joining a project. 
Study S21 analyzed the technical factors of past experience and social factors of past connections to understand onboarding in software projects.
\end{enumerate}

%\subsection{Threats to Validity}
\section{Threats to Validity}
%In this section, we discuss three key threats to the validity of the study.
%All threats are related to the review method that was used.

A key threat to the systematic mapping is the selection process, which was mostly carried out by the first author of this paper.
The initial round was mostly done using the abstract and titles.
To mitigate this bias, the next rounds of exclusions and inclusions were validated by two other co-authors. 
As shown in Figure \ref{fig:mappingoverview}, which follows the mapping guidelines \citep{Kitchenham2007}, five method papers and four survey papers that were initially discarded were later included in the final consolidation of the papers.

The second possible bias is the keywords that were used in the search string.
An alternative method includes performing a snowballing of references to get papers \citep{Jalali:2012:SLS:2372251.2372257}.
To mitigate this threat, we used reliable sources of only the premium conferences and recognized journals in software engineering (i.e., ranked as A to A++ only with impact factors higher than 2.63)\footnote{taken from \url{http://www.core.edu.au} rankings}.

The final threat to the study is the scope of exclusion and inclusion of the papers collected. 
%\myworries{(this is not true) First, our papers does not include other important studies that were not published from the nine venues (i.e., specialized conferences and workshops).
%We argue that our included papers represent the top-tier of mature technical contributions to software engineering.}
First, our scope does not include any papers published before 2013.
In line with our main goal, we would like to understand the more recent and applicable metrics that measure modern OSS projects.
Hence, the study of older papers is prone to reveal obsolete metrics that cannot be applied to current technology platforms (i.e., GitHub).
Second, as shown in Figure \ref{fig:term}, our search terms revolves around the eight inclusion keywords and fourteen exclusion terms.
A threat is that there could be other papers that were excluded due to this exclusion.
We are confident that this case is highly unlikely, as those papers would have a more narrow focus (i.e., human aspects in cloud technologies).
Furthermore, since the SE human capital is a generalized abstract of human capital, we assume that the metrics used in those specialized papers would not be useful.

\section{Conclusion}
\label{sec:conclusion}
In this paper, we use a systematic mapping to understand human capital, especially in the field of software engineering. 
We carried out a systematic mapping from five top journal articles and four top international conferences published between 2013 and 2017. 
Based on the economic Global Human Capital Index (GHCI) framework, we then identified and classified 78 studies into four dimensions: \textit{capacity} for skill attainment, \textit{deployment} for a workforce, \textit{development} for upskilling and reskilling, and \textit{know-how} for specialized skills.

The key outcome of this mapping study is a set of indicators for future constructing a software engineering (SE) HCI.
Much like the GHCI, SE-HCI can be used to understand how projects develop their human capital and be used as an important determinant of their long-term success than virtually any other factors.
We envision that HCI as a ranking of software projects to capture the full human capital potential profile.
For future work, we aim to implement the SE-HCI as a tool to assess progress within projects and point to opportunities for cross-project learning and exchange across projects.

%\begin{acknowledgements}
%Thank you
%\end{acknowledgements}

% BibTeX users please use one of
%\bibliographystyle{spbasic}      % basic style, author-year citations
%\bibliographystyle{spmpsci}      % mathematics and physical sciences
%\bibliographystyle{spphys}       % APS-like style for physics
%\bibliography{}   % name your BibTeX data base

% Non-BibTeX users please use
%\begin{thebibliography}{}
%
% and use \bibitem to create references. Consult the Instructions
% for authors for reference list style.
%
%\bibitem{RefJ}
% Format for Journal Reference
%Author, Article title, Journal, Volume, page numbers (year)
% Format for books
%\end{thebibliography}

%\begin{comment}
\appendix
\section{References of Collected Papers}
\label{sec:papers}

\paragraph{S01}\cite{SMS01}
Abdul~Rehman Gilal, Jafreezal Jaafar, Mazni Omar, Shuib Basri, and Ahmad Waqas.
\newblock {A rule-based model for software development team composition: Team
  leader role with personality types and gender classification}.
\newblock {\em Information and Software Technology}, 74:105--113, 2016.
\\ \textbf{Topics: team, personality, performance, gender}
\paragraph{S02}\cite{SMS02}
Marc Palyart, Gail~C. Murphy, and Vaden Masrani.
\newblock {A Study of Social Interactions in Open Source Component Use}.
\newblock {\em IEEE Transactions on Software Engineering}, 5589(c):1--1, 2017.
\\ \textbf{Topics: social interaction, community, OSS}
\paragraph{S03}\cite{SMS03}
Muneera Bano and Didar Zowghi.
\newblock {A systematic review on the relationship between user involvement and
  system success}.
\newblock {\em Information and Software Technology}, 58:148--169, 2015.
\\ \textbf{Topics: user, success, challenge, SLR}
\paragraph{S04}\cite{SMS04}
Elizabeth Bjarnason, Kari Smolander, Emelie Engstr{\"{o}}m, and Per Runeson.
\newblock {A theory of distances in software engineering}.
\newblock {\em Information and Software Technology}, 70:204--219, 2016.
\\ \textbf{Topics: distance, challenge, practice} 
\paragraph{S05}\cite{SMS05}
Georgios Gousios, Martin Pinzger, and Arie {Van Deursen}.
\newblock {An Exploratory Study of the Pull-Based Software Development Model}.
\newblock {\em Proceedings of the 36th International Conference on Software Engineering}, 345--355, 2014.
\\ \textbf{Topics: GitHub, community, pull request} 
\paragraph{S06}\cite{SMS06}
Aurora Vizca{\'{i}}no, F{\'{e}}lix Garc{\'{i}}a, Jos{\'{e}}~Carlos Villar,
  Mario Piattini, and Javier Portillo.
\newblock {Applying Q-methodology to analyse the success factors in GSD}.
\newblock {\em Information and Software Technology}, 55:1200--1211, 2013.
\\ \textbf{Topics: success, GSD, challenge, knowledge} 
\paragraph{S07}\cite{SMS07}
Klaus-Benedikt Schultis, Christoph Elsner, and Daniel Lohmann.
\newblock {Architecture Challenges for Internal Software Ecosystems: A
  Large-Scale Industry Case Study}.
\newblock {\em Proceedings of the 22nd ACM SIGSOFT International Symposium on Foundations of Software Engineering}, 542--552, 2014.
\\ \textbf{Topics: ecosystem, industry, challenge} 
\paragraph{S08}\cite{SMS08}
Marco Ortu, Bram Adams, Giuseppe Destefanis, Parastou Tourani, Michele
  Marchesi, and Roberto Tonelli.
\newblock {Are Bullies more Productive? Empirical Study of Affectiveness vs.
  Issue Fixing Time}.
\newblock {\em Proceedings of the 12th IEEE/ACM Working Conference on Mining Software Repositories},
303--313, 2015.
\\ \textbf{Topics: emotion, productivity, developer} 
\paragraph{S09}\cite{SMS09}
Silvia~T Acu{\~{n}}a, Marta~N G{\'{o}}mez, Jo~E Hannay, Natalia Juristo, and
  Dietmar Pfahl.
\newblock {Are team personality and climate related to satisfaction and
  software quality? Aggregating results from a twice replicated experiment}.
\newblock {\em Information and Software Technology}, 57:141--156, 2015.
\\ \textbf{Topics: team, personality, software quality, satisfaction} 
\paragraph{S10}\cite{SMS10}
Prem Devanbu, Thomas Zimmermann, and Christian Bird.
\newblock {Belief \& Evidence in Empirical Software Engineering}.
\newblock {\em Proceedings of the 38th IEEE/ACM International Conference on Software Engineering}, 108--119, 2016.
\\ \textbf{Topics: developer, practice, evidence} 
\paragraph{S11}\cite{SMS11}
Krzysztof Wnuk, Per Runeson, Matilda Lantz, and Oskar Weijden.
\newblock {Bridges and barriers to hardware-dependent software ecosystem
  participation - A case study}.
\newblock {\em Information and Software Technology}, 56(11):1493--1507, 2014.
\\ \textbf{Topics: barrier, ecosystem, communication} 
\paragraph{S12}\cite{SMS12}
Qi~Xuan and Vladimir Filkov.
\newblock {Building It Together: Synchronous Development in OSS}.
\newblock {\em Proceedings of the 36th International Conference on Software Engineering},
222--233, 2014.
\\ \textbf{Topics: OSS, productivity, communication} 
\paragraph{S13}\cite{SMS13}
Mahmood Niazi, Sajjad Mahmood, Mohammad Alshayeb, Mohammed~Rehan Riaz, Kanaan
  Faisal, Narciso Cerpa, Siffat~Ullah Khan, and Ita Richardson.
\newblock {Challenges of project management in global software development: A
  client-vendor analysis}.
\newblock {\em Information and Software Technology}, 80:1--19, 2016.
\\ \textbf{Topics: challenge, GSD, SLR} 
\paragraph{S14}\cite{SMS14}
Hideaki Hata, Taiki Todo, Saya Onoue, and Kenichi Matsumoto.
\newblock {Characteristics of Sustainable OSS Projects : A Theoretical and Empirical Study}.
\newblock {\em Proceedings of the 8th International Workshop on Cooperative and Human Aspects of Software Engineering}, 15--21, 2015.
\\ \textbf{Topics: OSS, community, GitHub} 
\paragraph{S15}\cite{SMS15}
Mitchell Joblin, Sven Apel, Claus Hunsen, and Wolfgang Mauerer.
\newblock {Classifying Developers into Core and Peripheral: An Empirical Study on Count and Network Metrics}.
\newblock {\em Proceedings of the 39th IEEE/ACM International Conference on Software Engineering},
164--174, 2017.
\\ \textbf{Topics: developer, network, knowledge} 
\paragraph{S16}\cite{SMS16}
Sherlock~A Licorish and Stephen~G Macdonell.
\newblock {Communication and personality profiles of global software developers}.
\newblock {\em Information and Software Technology}, 64:113--131, 2015.
\\ \textbf{Topics: communication, personality, GSD} 
\paragraph{S17}\cite{SMS17}
Anja Guzzi, Alberto Bacchelli, Michele Lanza, Martin Pinzger, and Arie {Van Deursen}.
\newblock {Communication in open source software development mailing lists}.
\newblock {\em Proceedings of the 10th IEEE International Working Conference on Mining Software Repositories}, 277--286, 2013.
\\ \textbf{Topics: communication, OSS, team} 
\paragraph{S18}\cite{SMS18}
G~R Bergersen, D~I~K Sjoberg, and T~Dyba.
\newblock {Construction and Validation of an Instrument for Measuring
  Programming Skill}.
\newblock {\em IEEE Transactions on Software Engineering}, 40(12):1163--1184, 2014.
\\ \textbf{Topics: success, practice, knowledge, performance} 
\paragraph{S19}\cite{SMS19}
Marcelo Cataldo and James~D. Herbsleb.
\newblock {Coordination breakdowns and their impact on development productivity and software failures}.
\newblock {\em IEEE Transactions on Software Engineering}, 39(3):343--360, 2013.
\\ \textbf{Topics: productivity, success, team, software quality} 
\paragraph{S20}\cite{SMS20}
Pamela Bhattacharya, Iulian Neamtiu, and Michalis Faloutsos.
\newblock {Determining developers' expertise and role: A graph hierarchy-based
  approach}.
\newblock {\em Proceedings of the 30th International Conference on Software Maintenance and Evolution}, 11--20, 2014.
\\ \textbf{Topics: developer, expertise, community} 
\paragraph{S21}\cite{SMS21}
Casey Casalnuovo, Bogdan Vasilescu, Premkumar Devanbu, and Vladimir Filkov.
\newblock Developer onboarding in github: The role of prior social links and
  language experience.
\newblock {\em Proceedings of the 10th Joint Meeting on Foundations of Software Engineering}, 817--828, 2015.
\\ \textbf{Topics: GitHub, team, productivity} 
\paragraph{S22}\cite{SMS22}
Kathryn~T Stolee, Sebastian Elbaum, and Anita Sarma.
\newblock {Discovering how end-user programmers and their communities use
  public repositories: A study on Yahoo! Pipes}.
\newblock {\em Information and Software Technology}, 55:1289--1303, 2013.
\\ \textbf{Topics: community, user, challenge, popularity} 
\paragraph{S23}\cite{SMS23}
Daryl Posnett, Raissa {D 'souza}, Premkumar Devanbu, and Vladimir Filkov.
\newblock {Dual Ecological Measures of Focus in Software Development}.
\newblock {\em Proceedings of the 35th International Conference on Software Engineering},
452--461, 2013.
\\ \textbf{Topics: practice, software quality, network} 
\paragraph{S24}\cite{SMS24}
Stefan Cedergren and Stig Larsson.
\newblock {Evaluating performance in the development of software-intensive
  products}.
\newblock {\em Information and Software Technology}, 56:516--526, 2014.
\\ \textbf{Topics: performance, practice, organization} 
\paragraph{S25}\cite{SMS25}
Tobias Baum, Olga Liskin, Kai Niklas, and Kurt Schneider.
\newblock {Factors Influencing Code Review Processes in Industry}.
\newblock {\em Proceedings of the 24th ACM SIGSOFT International Symposium on Foundations of Software Engineering}, 85--96, 2016.
\\ \textbf{Topics: code review, industry, OSS} 
\paragraph{S26}\cite{SMS26}
David Piorkowski, Austin~Z Henley, Tahmid Nabi, Scott~D Fleming, Christopher
  Scaffidi, and Margaret Burnett.
\newblock {Foraging and Navigations, Fundamentally: Developers' Predictions of
  Value and Cost}.
\newblock {\em Proceedings of the 24th ACM SIGSOFT International Symposium
  on Foundations of Software Engineering}, 97--108, 2016.
\\ \textbf{Topics: developer, IFT} 
\paragraph{S27}\cite{SMS27}
Mitchell Joblin, Wolfgang Mauerer, Sven Apel, Janet Siegmund, and Dirk Riehle.
\newblock {From Developer Networks to Verified Communities: A Fine-Grained
  Approach}.
\newblock {\em Proceedings of the 37th IEEE/ACM International Conference on Software
  Engineering}, 563--573, 2015.
\\ \textbf{Topics: network, community, software quality, productivity} 
\paragraph{S28}\cite{SMS28}
Anna Filippova, Erik Trainer, and James~D Herbsleb.
\newblock {From Diversity by Numbers to Diversity as Process: Supporting
  Inclusiveness in Software Development Teams with Brainstorming}.
\newblock {\em Proceedings of the 39th IEEE/ACM International Conference on Software Engineering},
152--163, 2017.
\\ \textbf{Topics: process, team, developer, satisfaction} 
\paragraph{S29}\cite{SMS29}
Sebastiano Panichella, Gabriele Bavota, Massimiliano {Di Penta}, Gerardo
  Canfora, and Giuliano Antoniol.
\newblock {How developers' collaborations identified from different sources
  tell us about code changes}.
\newblock {\em Proceedings of the 30th International Conference on Software
  Maintenance and Evolution}, 251--260, 2014.
\\ \textbf{Topics: developer, network, communication, OSS} 
\paragraph{S30}\cite{SMS30}
Wanwangying Ma, Lin Chen, Xiangyu Zhang, Yuming Zhou, and Baowen Xu.
\newblock {How do Developers Fix Cross-project Correlated Bugs? A case study on
  the GitHub scientific Python ecosystem}.
\newblock {\em Proceedings of the 39th IEEE/ACM International Conference on Software Engineering},
381--392, 2017.
\\ \textbf{Topics: GitHub, ecosystem, challenge, practice} 
\paragraph{S31}\cite{SMS31}
Jefferson~O. Silva, Igor Wiese, Daniel German, Igor Steinmacher, and Marco~A
  Gerosa.
\newblock {How Long and How Much : What to Expect from Summer of Code
  Participants ?}
\newblock {\em Proceedings of the 33rd International Conference on Software
  Maintenance and Evolution}, 69--79, 2017.
\\ \textbf{Topics: OSS, community, success, participation, experience} 
\paragraph{S32}\cite{SMS32}
Jason Tsay, Laura Dabbish, and James Herbsleb.
\newblock {Influence of Social and Technical Factors for Evaluating
  Contribution in GitHub}.
\newblock {\em Proceedings of the 36th International Conference on Software
  Engineering}, 356--366, 2014.
\\ \textbf{Topics: GitHub, OSS, practice, pull request} 
\paragraph{S33}\cite{SMS33}
Santiago Gala-P{\'{e}}rez, Gregorio Robles, Jes{\'{u}}s~M
  Gonz{\'{a}}lez-Barahona, and Israel Herraiz.
\newblock {Intensive Metrics for the Study of the Evolution of Open Source
  Projects: Case Studies from Apache Software Foundation Projects}.
\newblock {\em Proceedings of the 10th Working Conference on Mining Software Repositories},
159--168, 2013.
\\ \textbf{Topics: evolution, OSS, popularity} 
\paragraph{S34}\cite{SMS34}
Saya Onoue, Hideaki Hata, Akito Monden, and Kenichi Matsumoto.
\newblock {Investigating and projecting population structures in open source
  software projects: A case study of projects in GitHub}.
\newblock {\em IEICE Transactions on Information and Systems}, E99D(5):1304--1315, 2016.
\\ \textbf{Topics: OSS, GitHub, community, population} 
\paragraph{S35}\cite{SMS35}
Jason Tsay, Laura Dabbish, and James Herbsleb.
\newblock {Let's Talk About It: Evaluating Contributions through Discussion in
  GitHub}.
\newblock {\em Proceedings of the 22nd ACM SIGSOFT International Symposium on
  Foundations of Software Engineering}, 144--154, 2014.
\\ \textbf{Topics: GitHub, OSS, pull request, community} 
\paragraph{S36}\cite{SMS36}
Slinger Jansen.
\newblock {Measuring the health of open source software ecosystems: Beyond the
  scope of project health}.
\newblock {\em Information and Software Technology}, 56(11):1508--1519, 2014.
\\ \textbf{Topics: OSS, ecosystem, user, health, participation} 
\paragraph{S37}\cite{SMS37}
Eirini Kalliamvakou, Daniela Damian, Kelly Blincoe, Leif Singer, and Daniel~M
  German.
\newblock {Open Source-Style Collaborative Development Practices in Commercial
  Projects Using GitHub}.
\newblock {\em Proceedings of the 37th IEEE/ACM International Conference on Software Engineering}, 574--585, 2015.
\\ \textbf{Topics: OSS, practice, GitHub, popularity, pull request} 
\paragraph{S38}\cite{SMS38}
Denae Ford, Justin Smith, Philip~J Guo, and Chris Parnin.
\newblock {Paradise Unplugged: Identifying Barriers for Female Participation on
  Stack Overflow}.
\newblock {\em Proceedings of the 24th ACM SIGSOFT International Symposium
  on Foundations of Software Engineering}, 846--857, 2016.
\\ \textbf{Topics: barrier, community, expertise} 
\paragraph{S39}\cite{SMS39}
Xin Xia, David Lo, Lingfeng Bao, Abhishek Sharma, and Shanping Li.
\newblock {Personality and Project Success : Insights from a Large-Scale Study
  with Professionals}.
\newblock {\em Proceedings of the 33rd IEEE International Conference on Software Maintenance and
  Evolution}, 318--328, 2017.
\\ \textbf{Topics: personality, success, emotion} 
\paragraph{S40}\cite{SMS40}
Peter~C Rigby, Yue~Cai Zhu, Samuel~M Donadelli, and Audris Mockus.
\newblock {Quantifying and Mitigating Turnover-Induced Knowledge Loss: Case
  Studies of Chrome and a project at Avaya}.
\newblock {\em Proceedings of the 38th IEEE/ACM International Conference on Software Engineering},
1006--1016, 2016.
\\ \textbf{Topics: turnover, knowledge, productivity} 
\paragraph{S41}\cite{SMS41}
J~M Verner, O~P Brereton, B~A Kitchenham, M~Turner, and M~Niazi.
\newblock {Risks and risk mitigation in global software development: A tertiary
  study}.
\newblock {\em Information and Software Technology}, 56(1):54--78, 2014.
\\ \textbf{Topics: risk, GSD, SLR} 
\paragraph{S42}\cite{SMS42}
Mohammad Gharehyazie and Vladimir Filkov.
\newblock {Tracing distributed collaborative development in apache software
  foundation projects}.
\newblock {\em Empirical Software Engineering}, 22(4):1795--1830, 2017.
\\ \textbf{Topics: developer, OSS, team, productivity} 
\paragraph{S43}\cite{SMS43}
Ricardo~M Czekster, Paulo Fernandes, Lucelene Lopes, Afonso Sales, Alan~R
  Santos, and Thais Webber.
\newblock {Stochastic Performance Analysis of Global Software Development
  Teams}.
\newblock {\em ACM Trans. Softw. Eng. Methodol. Article}, 25(26), 2016.
\\ \textbf{Topics: performance, GSD, productivity} 
\paragraph{S44}\cite{SMS44}
Sebastian~C M{\"{u}}ller and Thomas Fritz.
\newblock {Stuck and Frustrated or In Flow and Happy: Sensing Developers'
  Emotions and Progress}.
\newblock {\em Proceedings of the 37th IEEE/ACM International Conference on Software
  Engineering}, 688--699, 2015.
\\ \textbf{Topics: emotion, productivity} 
\paragraph{S45}\cite{SMS45}
Julia Rubin and Martin Rinard.
\newblock {The Challenges of Staying Together While Moving Fast: An Exploratory
  Study}.
\newblock {\em Proceedings of the 38th IEEE/ACM International Conference on Software Engineering},
982--993, 2016.
\\ \textbf{Topics: challenge, satisfaction, community} 
\paragraph{S46}\cite{SMS46}
Paul Ralph and Paul Kelly.
\newblock {The Dimensions of Software Engineering Success}.
\newblock {\em Proceedings of the 36th International Conference on Software
  Engineering}, 24--35, 2014.
\\ \textbf{Topics: success, practice, satisfaction, performance} 
\paragraph{S47}\cite{SMS47}
Arjumand~Bano Soomro, Norsaremah Salleh, Emilia Mendes, John Grundy, Giles
  Burch, and Azlin Nordin.
\newblock {The effect of software engineers' personality traits on team climate
  and performance: A Systematic Literature Review}.
\newblock {\em Information and Software Technology}, 73:52--65, 2016.
\\ \textbf{Topics: personality, performance, success, team} 
\paragraph{S48}\cite{SMS48}
Anh Nguyen-Duc, Daniela~S Cruzes, and Reidar Conradi.
\newblock {The impact of global dispersion on coordination, team performance
  and software quality -- A systematic literature review}.
\newblock {\em Information and Software Technology}, 57:277--294, 2015.
\\ \textbf{Topics: GSD, SLR, performance, software quality} 
\paragraph{S49}\cite{SMS49}
Youngsoo Kim and Lingxiao Jiang.
\newblock {The Learning Curves in Open-Source Software (OSS) Development Network}.
\newblock {\em Proceedings of the 6th International Conference on Electronic Commerce},
41--48, 2014.
\\ \textbf{Topics: OSS, network, developer, learning} 
\paragraph{S50}\cite{SMS50}
Daniela Damian, Remko Helms, Irwin Kwan, Sabrina Marczak, and Benjamin
  Koelewijn.
\newblock {The Role of Domain Knowledge and Cross-Functional Communication in
  Socio-Technical Coordination}.
\newblock {\em Proceedings of the 35th International Conference on Software Engineering},
442--451, 2013.
\\ \textbf{Topics: knowledge, communication, team} 
\paragraph{S51}\cite{SMS51}
K\i van{\c{c}} Mu{\c s}lu, Christian Bird, Nachiappan Nagappan, and
  Jacek Czerwonka.
\newblock {Transition from Centralized to Decentralized Version Control
  Systems: A Case Study on Reasons, Barriers, and Outcomes}.
\newblock {\em Proceedings of the 36th IEEE/ACM International Conference on Software
  Engineering}, 334--344, 2014.
\\ \textbf{Topics: barrier, developer, team} 
\paragraph{S52}\cite{SMS52}
Shaun Phillips, Thomas Zimmermann, and Christian Bird.
\newblock {Understanding and Improving Software Build Teams}.
\newblock {\em Proceedings of the 36th International Conference on Software
  Engineering}, 735--744, 2014.
\\ \textbf{Topics: team, challenge, knowledge, practice} 
\paragraph{S53}\cite{SMS53}
Jing Jiang, David Lo, Xinyu Ma, Fuli Feng, and Li~Zhang.
\newblock {Understanding inactive yet available assignees in GitHub}.
\newblock {\em Information and Software Technology}, 91:44--55, 2017.
\\ \textbf{Topics: GitHub, OSS, pull request, available} 
\paragraph{S54}\cite{SMS54}
Sherlock~A Licorish and Stephen~G Macdonell.
\newblock {Understanding the attitudes, knowledge sharing behaviors and task
  performance of core developers: A longitudinal study}.
\newblock {\em Information and Software Technology}, 56:1578--1596, 2017.
\\ \textbf{Topics: knowledge, performance, success} 
\paragraph{S55}\cite{SMS55}
Hudson Borges, Andre Hora, and Marco~Tulio Valente.
\newblock {Understanding the factors that impact the popularity of GitHub
  repositories}.
\newblock {\em Proceedings of the 32nd IEEE International Conference on Software
  Maintenance and Evolution}, 334--344, 2016.
\\ \textbf{Topics: popularity, GitHub, OSS, user} 
\paragraph{S56}\cite{SMS56}
Amanda Lee, Jeffrey~C Carver, and Amiangshu Bosu.
\newblock {Understanding the Impressions, Motivations, and Barriers of One Time
  Code Contributors to FLOSS Projects: A Survey}.
\newblock {\em Proceedings of the 39th IEEE/ACM International Conference on Software Engineering},
187--197, 2017.
\\ \textbf{Topics: barrier, success, OSS} 
\paragraph{S57}\cite{SMS57}
Kelly Blincoe, Jyoti Sheoran, Sean Goggins, Eva Petakovic, and Daniela Damian.
\newblock {Understanding the popular users: Following, affiliation influence
  and leadership on GitHub}.
\newblock {\em Information and Software Technology}, 70:30--39, 2016.
\\ \textbf{Topics: user, OSS, GitHub, popularity} 
\paragraph{S58}\cite{SMS58}
Mansooreh Zahedi and Muhammad~Ali Babar.
\newblock {Why does site visit matter in global software development: A knowle
  dge-base d perspective}.
\newblock {\em Information and Software Technology}, 80:36--56, 2016.
\\ \textbf{Topics: GSD, knowledge, practice} 
\paragraph{S59}\cite{SMS59}
Mathieu Lavall{\'{e}}e and Pierre~N Robillard.
\newblock {Why Good Developers Write Bad Code: An Observational Case Study of
  the Impacts of Organizational Factors on Software Quality}.
\newblock {\em Proceedings of the 37th IEEE/ACM IEEE International Conference on Software
  Engineering}, 677--687, 2015.
\\ \textbf{Topics: software quality, GitHub, organization, structure} 
\paragraph{S60}\cite{SMS60}
Jailton Coelho and Marco~Tulio Valente.
\newblock {Why Modern Open Source Projects Fail}.
\newblock {\em Proceedings of the 11th Joint Meeting on Foundations of Software Engineering}, 186--196, 2017.
\\ \textbf{Topics: OSS, practice, success} 
\paragraph{S61}\cite{SMS61}
Georgios Gousios, Margaret-Anne Storey, and Alberto Bacchelli.
\newblock {Work Practices and Challenges in Pull-Based Development: The
  Contributor's Perspective}.
\newblock {\em Proceedings of the 38th IEEE/ACM International Conference on Software Engineering},
285--296, 2016.
\\ \textbf{Topics: challenge, OSS, practice, GitHub, pull request} 
\paragraph{S62}\cite{SMS62}
Georgios Gousios, Andy Zaidman, Margaret-Anne Storey, and Arie {Van Deursen}.
\newblock {Work Practices and Challenges in Pull-Based Development: The
  Integrator's Perspective}.
\newblock {\em Proceedings of the 37th IEEE/ACM International Conference on Software
  Engineering}, 358--368, 2015.
\\ \textbf{Topics: practice, challenge, pull request, participation} 
\paragraph{S63}\cite{SMS63}
Margaret~Anne Storey, Alexey Zagalsky, Fernando~Figueira Filho, Leif Singer,
  and Daniel~M. German.
\newblock {How Social and Communication Channels Shape and Challenge a
  Participatory Culture in Software Development}.
\newblock {\em IEEE Transactions on Software Engineering}, 43(2):185--204, 2017.
\\ \textbf{Topics: communication, developer, GitHub} 
\paragraph{S64}\cite{SMS64}
Bram Adams, Ryan Kavanagh, Ahmed~E. Hassan, and Daniel~M. German.
\newblock {An empirical study of integration activities in distributions of
  open source software}.
\newblock {\em Empirical Software Engineering}, 21(3):960--1001, 2016.
\\ \textbf{Topics: integration, OSS, reuse, practice} 
\paragraph{S65}\cite{SMS65}
Per Lenberg, Lars~G{\"{o}}ran {Wallgren Tengberg}, and Robert Feldt.
\newblock {An initial analysis of software engineers' attitudes towards
  organizational change}.
\newblock {\em Empirical Software Engineering}, 22(4):2179--2205, 2017.
\\ \textbf{Topics: attitude, organization, process} 
\paragraph{S66}\cite{SMS66}
David Kavaler and Vladimir Filkov.
\newblock {Stochastic actor-oriented modeling for studying homophily and social
  influence in OSS projects}.
\newblock {\em Empirical Software Engineering}, 22(1):407--435, 2017.
\\ \textbf{Topics: OSS, network, productivity} 
\paragraph{S67}\cite{SMS67}
Yi~Wang and David Redmiles.
\newblock {Cheap talk, cooperation, and trust in global software engineering}.
\newblock {\em Empirical Software Engineering}, 21(6):2233--2267, 2016.
\\ \textbf{Topics: communication, GSD, EGT} 
\paragraph{S68}\cite{SMS68}
Ulrike Abelein, Barbara Paech, Daniela Damian, U~Abelein, and B~Paech.
\newblock {Understanding the Influence of User Participation and Involvement on
  System Success -- a Systematic Mapping Study}.
\newblock {\em Empirical Software Eng}, 20:28--81, 2015.
\\ \textbf{Topics: user, success, SLR, participation} 
\paragraph{S69}\cite{SMS69}
Mohammad Gharehyazie, Daryl Posnett, Bogdan Vasilescu, Vladimir Filkov,
  Yann-Ga{\"{e}}l Gu{\'{e}}h{\'{e}}neuc, Tom Mens, M~Gharehyazie, D~Posnett,
  V~Filkov, and B~Vasilescu.
\newblock {Developer initiation and social interactions in OSS: A case study of
  the Apache Software Foundation}.
\newblock {\em Empirical Software Eng}, 20:1318--1353, 2015.
\\ \textbf{Topics: developer, social interaction, OSS, communication} 
\paragraph{S70}\cite{SMS70}
Mitchell Joblin, Sven Apel, and Wolfgang Mauerer.
\newblock {Evolutionary trends of developer coordination: a network approach}.
\newblock {\em Empirical Software Engineerings}, 22(4):2050--2094, 2017.
\\ \textbf{Topics: developer, network, evolution} 
\paragraph{S71}\cite{SMS71}
Ingo Scholtes, Pavlin Mavrodiev, and Frank Schweitzer.
\newblock {From Aristotle to Ringelmann: a large-scale analysis of team
  productivity and coordination in Open Source Software projects}.
\newblock {\em Empirical Software Engineering}, 21(2):642--683, 2016.
\\ \textbf{Topics: productivity, OSS, team} 
\paragraph{S72}\cite{SMS72}
Alexey Zagalsky, Daniel~M. German, Margaret~Anne Storey, Carlos~G{\'{o}}mez
  Teshima, and Germ{\'{a}}n Poo-Caama{\~{n}}o.
\newblock {How the R community creates and curates knowledge: an extended study
  of stack overflow and mailing lists}.
\newblock {\em Empirical Software Engineering}, pages 1--34, 2017.
\\ \textbf{Topics: community, knowledge, stack overflow} 
\paragraph{S73}\cite{SMS73}
Olga Baysal, Oleksii Kononenko, Reid Holmes, and Michael~W. Godfrey.
\newblock {Investigating technical and non-technical factors influencing modern
  code review}.
\newblock {\em Empirical Software Engineering}, 21(3):932--959, 2016.
\\ \textbf{Topics: code review, developer, patch submission} 
\paragraph{S74}\cite{SMS74}
{\"{O}}zlem Albayrak and Jeffrey~C Carver.
\newblock {Investigation of individual factors impacting the effectiveness of
  requirements inspections: a replicated experiment}.
\newblock {\em Empirical Software Engineering}, 2014.
\\ \textbf{Topics: requirement, developer, inspection, individual} 
\paragraph{S75}\cite{SMS75}
Nicolas Bettenburg, Ahmed~E Hassan, Bram Adams, Daniel~M German, N~Bettenburg,
  A~E Hassan, B~Adams, and D~M German.
\newblock {Management of community contributions A case study on the Android
  and Linux software ecosystems}.
\newblock {\em Empirical Software Engineering}, 2015.
\\ \textbf{Topics: community, OSS, practice, management} 
\paragraph{S76}\cite{SMS76}
Bogdan Vasilescu, Alexander Serebrenik, Mathieu Goeminne, Tom Mens,
  Margaret-Anne Storey, B~Vasilescu, A~Serebrenik,
  A~Serebrenik, M~Goeminne, and T~Mens.
\newblock {On the variation and specialisation of workload---A case study of
  the GNOME ecosystem community}.
\newblock {\em Empirical Software Eng}, 19:955--1008, 2014.
\\ \textbf{Topics: ecosystem, community, OSS} 
\paragraph{S77}\cite{SMS77}
Cristina Palomares, Carme Quer, and Xavier Franch.
\newblock {Requirements reuse and requirement patterns: a state of the practice
  survey}.
\newblock {\em Empirical Software Engineering}, pages 1--44, 2017.
\\ \textbf{Topics: requirement, practice, challenge} 
\paragraph{S78}\cite{SMS78}
Patanamon Thongtanunam, Shane McIntosh, Ahmed~E. Hassan, and Hajimu Iida.
\newblock {Review participation in modern code review: An empirical study of
  the android, Qt, and OpenStack projects}.
\newblock {\em Empirical Software Engineering}, 22(2):768--817, 2017.
\\ \textbf{Topics: software quality, practice, code review, participation} 
%\section*{Appendix}
%%%%%%%%%%%%%%%%%%%%%%%%%%%%%%%%%%%%%%%%%%%%%%%%%%%%%%%%%%%

%\end{comment}
%\end{landscape}
%%%%%%%%%%%%%%%%%%%%%%%%%%%%%%%%%%%%%%%%%%%%%%%%%%%%%%%%%%%
% BibTeX users please use one of
%\bibliographystyle{spbasic}      % basic style, author-year citations
%\bibliographystyle{spmpsci}      % mathematics and physical sciences
%\bibliographystyle{spphys}       % APS-like style for physics
%\bibliography{reference}

%\bibliographystyle{IEEEtran}
%\bibliography{reference}

\end{document}